# Combining Machine Learning Models with First-Principles High-Throughput Calculation to Accelerate the Search of Promising Thermoelectric Materials

Tao Fan*, Artem R. Oganov
Skolkovo Institute of Science and Technology, Bolshoy Boulevard 30, bld. 1, 121205 Moscow, Russia.

## ABSTRACT

Thermoelectric materials can achieve direct energy conversion between electricity and heat, thus can be applied to waste heat harvesting and solid-state cooling. The discovery of new thermoelectric materials is mainly based on experiments and first-principles calculations. However, these methods are usually expensive and time-consuming. Recently, the prediction of properties via machine learning has emerged as a popular method in materials science. Herein, we firstly did first-principles high-throughput calculations for a large number of chalcogenides and built a thermoelectric database containing 796 compounds. Many novel and promising thermoelectric materials were discovered. Then, we trained four ensemble learning models and two deep learning models to distinguish the promising thermoelectric materials from the others for n type and p type doping, respectively. All the presented models achieve classification accuracy higher than 85% and area under the curve (AUC) higher than 0.9. Especially, the M3GNet model for n type data achieve accuracy, precision and recall all higher than 90%. Our works demonstrate a very efficient way of combining machine learning prediction and first-principles high-throughput calculations together to accelerate the discovery of advanced thermoelectric materials.

## INTRODUCTION

Thermoelectric (TE) materials could play an important role in building clean and alternative energy sources due to their ability to realize the direct conversion between heat and electricity[1-3]. Thermoelectric devices have the characteristics of small size, no noise, and no pollution, thus they have wide application in space power, industrial waste heat harvesting, small and mobile refrigerators, and other fields[4-6]. The energy conversion efficiency of thermoelectric materials depends on the dimensionless figure of merit (*ZT*). *ZT* is defined as $ZT = \alpha^2\sigma T/(\kappa_e + \kappa_L)$, where $\alpha$ is the Seebeck coefficient, $\sigma$ is the electrical conductivity, $T$ is the absolute temperature, $\kappa_e$ is the electronic thermal conductivity, and $\kappa_L$ is the lattice thermal conductivity. Particularly, $\alpha^2\sigma$ is called the power factor (*PF*). In order to obtain a high *ZT*, both $\alpha$ and $\sigma$ must be maximized, while $\kappa_e$ and $\kappa_L$ need to be minimized. However, the interdependence of these parameters makes improving the *ZT* of a material a great challenge[7,8].

Traditional thermoelectric materials discovery has been led by experiments, while computations are becoming more and more important with the advance of theory and the increase of computing power[9-11]. First-principles methods, such as DFT, have been widely used in calculating the thermoelectric related properties[12-15]. However, full first-principles calculation of transport properties is usually computationally expensive. Thus, there are many simplified models being proposed to calculate electronic and phonon transport properties, leading to many interesting and important discoveries[16-20]. However, they all face the problem of accuracy-computational cost trade-off. Recently, machine learning (ML) has achieved much progress in both its theory and available models[21]. Data science and machine learning have become an integral part of natural sciences, thought as the fourth pillar in science, next to experiment, theory, and simulation[22-25]. ML algorithms find patterns in high-dimensional training data and build a mathematical model to make predictions or decisions without explicit human knowledge. This approach has been applied successfully to various materials science studies, such as structure predictions[26,27], constructing force field[28-30], and predictions of the static properties of materials[31-33]. A variety of ML algorithms provide an alternative to costly and complex DFT calculations, providing similarly accurate results in a fraction of time. For TE materials, ML-assisted research has also been conducted, focusing on predict the band structure, Seebeck coefficient, electrical conductivity, power factor, lattice thermal conductivity and figure of merit directly with carefully designed features or features learned by the model itself[34-39]. These works are based on supervised learning. Besides, there are some works using unsupervised ML method, which does not require well-labeled training data and can discover hidden patterns in the unlabeled datasets based on the input feature values. For example, Jia et al proposed a strategy to discover a series of promising half-Heusler thermoelectric materials through the iterative combination of unsupervised machine learning with labeled known half-Heusler thermoelectric materials[40].

*Corresponding author, E-mail: Tao.Fan@skoltech.ru

The training data for ML-assisted thermoelectric studies are usually from experiments. However, such data not only distributed in huge number of literatures, but also have varied synthesis conditions and measurement conditions. Even worse is the data in different literatures may conflict with each other. While first-principles calculations can guarantee the consistency of the data, the huge computational cost for high-throughput study limits the size of the obtainable data. In this work, we explore the possibility of using ML methods to speed up the first-principles high-throughput screening work. Using the software AICON developed in our group[41], we firstly built a database comprising of 796 chalcogenides with their n type and p type TE properties, including Seebeck coefficient, electrical conductivity, power factor, *etc*.. We found many novel and promising TE materials, and some of them, such as $Ge_5Te_4Se$, $KBiSe_2$, GeTe (*Pnma*) and $BaCu_2Te_2$, are even much better than those currently state-of-the-art TE materials. Then, we applied machine learning methods to the current datasets and trained four ensemble learning models — random forest (RF), gradient boosted decision tree (GBDT), adaptive boosting (AdaB), extreme gradient boosting (XGB)[42] — and two deep learning models — MatErials graph network (MEGNet)[43], materials three-body graph network (M3GNet)[44] — to identify the promising TE materials (label 1) from the others (label 0). The prediction on test sets show that all the trained models can achieve classification accuracy higher than 85%. Especially, the M3GNet model for n type data achieve accuracy, precision and recall all higher than 90%, showing great potential for pre-screening the TE materials without heavy calculations. Furthermore, the trained models were analyzed by SHAP method, revealing that these models truly capture the underline physics.

## METHODS

### First-principles calculation

All first-principles calculations were performed using the Vienna Ab initio Simulation Package (VASP) with the Perdew–Burke–Ernzerhof generalized gradient approximation (PBE–GGA) and the projector augmented wave (PAW) pseudo-potentials [45-47]. For structure relaxation, the plane wave kinetic energy cut-off was set to 600 eV and the Brillouin zone was sampled using Γ-centered meshes with the reciprocal-space resolution of $2\pi \times 0.03$ Å$^{-1}$. Kohn–Sham equations were solved self-consistently with the total energy tolerance of $10^{-7}$ eV/cell and structures were relaxed until the maximum force became smaller than $10^{-3}$ eV/Å. The dielectric constants were calculated using the DFPT [48], and the elastic constants were calculated using the finite difference method as implemented in VASP. To obtain the deformation potential constants of a compound, three band structure calculations were run: one at the equilibrium volume, the other two at volumes -0.1% and +0.1% with respect to the equilibrium one. The high symmetry path of band structure was generated by following the suggestion of this work[49].

After all necessary first-principles calculations finished, the resulting files were collected as input to software AICON to calculate TE transport properties. To enable high-throughput screening, the automated workflow control from structural relaxations, over band structures calculations *etc.*, to transport properties was also developed and implemented in AICON based on Materials Project high-throughput infrastructure[50-52]. The detail calculating process and settings for each step can be found in our previous work[41,53].

### Machine learning implementation

Feature engineering

Feature engineering is a crucial step in machine learning, as it directly impacts the performance of predictive models. The input features for a compound should be quick to compute and should capture all relevant features of that compound in a compact list of attributes. In this work, we designed a feature vector consisting of three different groups of descriptors:

*Composition*: the group of composition descriptors are similar to those used by Ward *et al.*[54]. It includes stoichiometric attributes, elemental property attributes, valence shell attributes, ionicity attributes. Different from the previous work, elemental property attributes are formed by mean, maximum, minimum, range, and mean absolute deviation of 23 different elemental properties of all atoms in a compound. The list of these 23 elemental properties can be found in supplementary information Table S1.

*Structure*: the group of structure descriptors have three different types, Voronoi tessellation of crystal structure (VORONOI) as used by Ward et al.[55], partial radial distribution function (PRDF) as used by Schutt et al.[56], and generalized radial distribution function plus bond order parameter (GRDF +BOP) as used by Seko et al.[57]. These three types structure descriptors were commonly adopted general purpose structure descriptors. Specifically, for PRDF descriptors we generated four feature vectors with different length by fixing the *cutoff* parameter as 20 Å and changing the *bin_size* parameter, they are PRDF_10 (*bin_size* = 0.1 Å), PRDF_16 (*bin_size* = 0.16 Å), PRDF_20 (*bin_size* = 0.2 Å), and PRDF_25 (*bin_size* = 0.25 Å).

*Band structure*: the group of band structure descriptors include five easily calculated band structure parameters, including band degeneracy *N*, conductivity effective mass $m_c^*$, density of states effective mass $m_d^*$, deformation potential constant Ξ, and band gap $E_g$. These parameters are also used by AICON to calculate the thermoelectric transport properties.

The composition and structure descriptors were generated by Matminer[58]. According to the involved structure descriptors, 6 different feature vectors were used in this work, namely PRDF_10, PRDF_16, PRDF_20, PRDF_25, VORONOI and GRDF_BOP. The python script used to generate these vectors are included in supplementary materials.

**Model selection**

In this work, four common ensemble machine learning algorithms, including random forest (RF), gradient boosted decision tree (GBDT), adaptive boosting (AdaB) as implemented in scikit-learn (*sklearn*) package[59], and extreme gradient boosting model as implemented in *XGBoost* package with *sklearn* interface[42], were used to train the classification models. Before entering the ML algorithm, the input feature vectors were firstly standardized to eliminate the influence of variance of each descriptor. A grid search method with 5-fold cross validation on training set as implemented in *sklearn* were used to optimize the hyperparameters in these models.

For deep learning, MEGNet and M3GNet models as implemented in Materials Graph Library (MGL) were used[43,44]. The binary cross entropy loss was used as loss function. We train all models for 550 epochs using the Adam optimizer and a batch size of 64. The initial learning rate is set to 0.001 and the *LinearLR* scheduler is used to adjust the learning rate per epoch. The final learning rate decays to 10% of the original value after 500 epochs, then keeps constant with another 50 epochs. During the optimization, the validation set' loss values were used to monitor the model's performance.

**Model evaluation**

The performance of ML models was evaluated with those metrics such as accuracy, precision, recall, F1-score, ROC curve and AUC. For a binary classification problem, all samples can be divided into four categories according to the combination of their real label and predicted label, as shown in Table 1,

Table 1. Confusion matrix for binary classification

| Reality | Prediction | |
|---|---|---|
| | Positive | Negative |
| Positive | TP | FN |
| Negative | FP | TN |

Then, the definition of those metrics are as follows:

$$\text{Accuracy} = \frac{\text{TP} + \text{TN}}{\text{TP} + \text{FN} + \text{FP} + \text{TN}}$$

$$\text{Precision} = \frac{\text{TP}}{\text{TP} + \text{FP}}$$

$$\text{Recall} = \text{TPR} = \frac{\text{TP}}{\text{TP} + \text{FN}}$$

$$\text{F1} = \frac{2 \times \text{Pre} \times \text{Rec}}{\text{Pre} + \text{Rec}}$$

$$\text{FPR} = \frac{\text{FP}}{\text{TN} + \text{FP}}$$

The Receiver Operating Characteristic (ROC) curve is the plot of the true positive rate (TPR) against the false positive rate (FPR) at each threshold setting. This means that the top left corner of the plot is the "ideal" point - a FPR of zero, and a TPR of one. AUC is the area under the ROC curve. Thus, for perfect classifier, AUC equals to 1. All of the above metrics were implemented in *sklearn.metrics*.

## RESULTS AND DISCUSSION

### Thermoelectric dataset

All structures were extracted from the Materials Project database[60] with five searching criteria: (1) S, Se and Te as anions; (2) the band gap should be larger than 0 eV but smaller than 1.2 eV, since good TE materials are usually narrow gap semiconductors; (3) the energy above the convex hull line should be less than 0.1 eV per atom to ensure the structure is thermodynamically stable or at least potentially synthesizable in experimental conditions; (4) belonging to cubic, tetragonal and orthorhombic crystal system; (5) nonferromagnetic phase. These criteria resulted in over 1000 entries out of the database. Then, calculations of electronic transport properties for these compounds were carried out. Finally, 796 compounds have finished the complete process of such calculations. Among them, 752 items of n type and 757 items of p type. Other structures could fail because of various reasons. For example, since GGA is known to underestimate the band gap, the calculated band gaps of some structures with very small gap values could be zero. In addition, the eigenvalues of the elastic constant matrix of some compounds have negative values. Such structures were just discarded.

Fig. 1 shows the maximum power factor with respect to the carrier concentration reaching this maximum value for n and p type compounds respectively in the temperature range from 300 K to 1000 K. The maximum power factors for both n type and p type compounds distribute across three order of magnitude and most of them are within the range of 1 $\mu W \cdot cm^{-1} \cdot K^{-2}$ — 10 $\mu W \cdot cm^{-1} \cdot K^{-2}$. Some representative compounds are marked in the pictures. Among these compounds, PbTe, PbS (cubic), PbSe (cubic), GeTe (cubic), SnTe, SnSe are already well-known TE materials and have high power factor values according to our calculation, which validates our methods. In our previous work, we have introduced those promising TE materials with cubic symmetry[53]. In supplementary information Table S2 and S3, we list top 50 non-cubic TE materials found in this work for n and p type doping, respectively. All of them have their $PF_{\max}$ larger than 10 μW·cm$^{-1}$·K$^{-2}$, similar to those well-known TE materials. The tables also list the band structure parameters for each compound. Generally, they all have large band degeneracy $N$ and small conductivity effective mass $m_c^*$. Small $m_c^*$ is beneficial for carrier mobility, while large $N$, which means there are many carrier pockets involved in the transport, can make DOS effective mass $m_d^*$ large, thus is beneficial for Seebeck coefficient. Some of the listed compounds, such as $Ge_5Te_4Se$, $KBiSe_2$, TbAsSe, DyAsSe, YAsSe, PbS (*Cmcm*), PbSe (*Cmcm*), GeTe (*Pnma*) etc., have high $PF_{\max}$ for both n and p type doping, which is good for building TE devices.

We picked up several interesting compounds and further calculated their lattice thermal conductivity $\kappa_L$, thus figure of merit $ZT$ using AICON[41]. For example, SnSe is a famous TE materials and its single crystal has record high $ZT$ 2.6 around 900 K[64]. SnSe has two phase: the low temperature $\alpha$ phase with space group *Pnma* and high temperature $\beta$ phase with space group *Cmcm*. In Table S2 and S3, there are several compounds which are similar to SnSe. Among them, the structures of GeTe, $SnPbS_2$ and GeSe are the same as $\alpha$-SnSe, while the structures of PbS and PbSe are the same as $\beta$-SnSe. The crystal structures of the above compounds are shown in Fig. 2. Fig. 3 shows the $PF$ of GeTe, $SnPbS_2$ and GeSe as a function of temperature and carrier concentration. Our calculation suggests that $SnPbS_2$ and GeSe are promising n type TE materials while GeTe is good for both n and p type. Fig. 4 shows the $\kappa_L$ of these three compounds. Similar with that of SnSe, the $\kappa_L$ of these compounds are very low within the calculated temperature range. Therefore, their $ZT$ values can exceed 1 in a wide range of temperatures and carrier concentrations, as shown in Fig. 5. Here we need to mention that AICON's lattice thermal conductivity model tends to underestimate the $\kappa_L$ value of those compounds with strong anharmonicity. Thus, the $ZT$ values of these compounds are overestimated much. Still, the values can be a sign of great potential of these compounds. Fig. S1, Fig S2 and Fig. S3 in supplementary information show the detail $PF$, $\kappa_L$ and $ZT$ values of *Cmcm* PbS and PbSe. These two compounds are excellent n and p type TE materials, their $ZT$ values could be higher than 1 in a wide range of temperatures and carrier concentrations. Another interesting compound we want to introduce is $BaCu_2Te_2$. This compound has high $PF$ for n type doping. The crystal structure of $BaCu_2Te_2$ is shown in Fig. S4. The Cu atom connects with four nearest neighboring Te atoms in a tetrahedral coordination. If viewing the structure along the *a* axis, the Cu and Te atoms form distorted rings and the hollow extends along the *a* axis, while the Ba atoms are distributed inside the hollow space. This structure is beneficial for impeding the transport of phonons, and thus is expected to have low lattice thermal conductivity. According to our calculation, it truly has very low $\kappa_L$ values, together with high $PF$ values, the $ZT$ of this compound are very large in a wide range of temperatures and carrier concentrations (see Fig. S5). We plan to release the complete database in the near future.

Although our high-throughput framework AICON is quite fast for transport properties calculations, it still has space to improve. The overall calculation process involves several different types of first-principles calculations. Some of these calculations, such as elastic constant and dielectric constant, are quite time-consuming. Since good thermoelectric materials are always minority, most of the computing resources are consumed on unpromising materials. If one could evaluate whether a compound was good or not good thermoelectric materials before doing computationally demanding first-principles calculations, it would save a lot of time and computing resources. The simplest idea is training a classification model to distinguish good and not good thermoelectric materials. At the first step, we need to label each sample in the dataset as positive (label 1) or negative (label 0). The value of maximum power factor $PF_{\max}$ as shown in Fig. 1 is a good index to be used to split the dataset. Here we used $PF_{\max}$ = 5 μW·cm$^{-1}$·K$^{-2}$ as the dividing line (the red line in Fig. 1). The compounds above the line were labeled as 1, while those under the line were labeled as 0. The explanation for using 5 μW·cm$^{-1}$·K$^{-2}$ as the boundary is in the supplementary information. According to such split, for n type data, #positive:#negative = 308:444, while for p type data, #positive:#negative = 244:513.

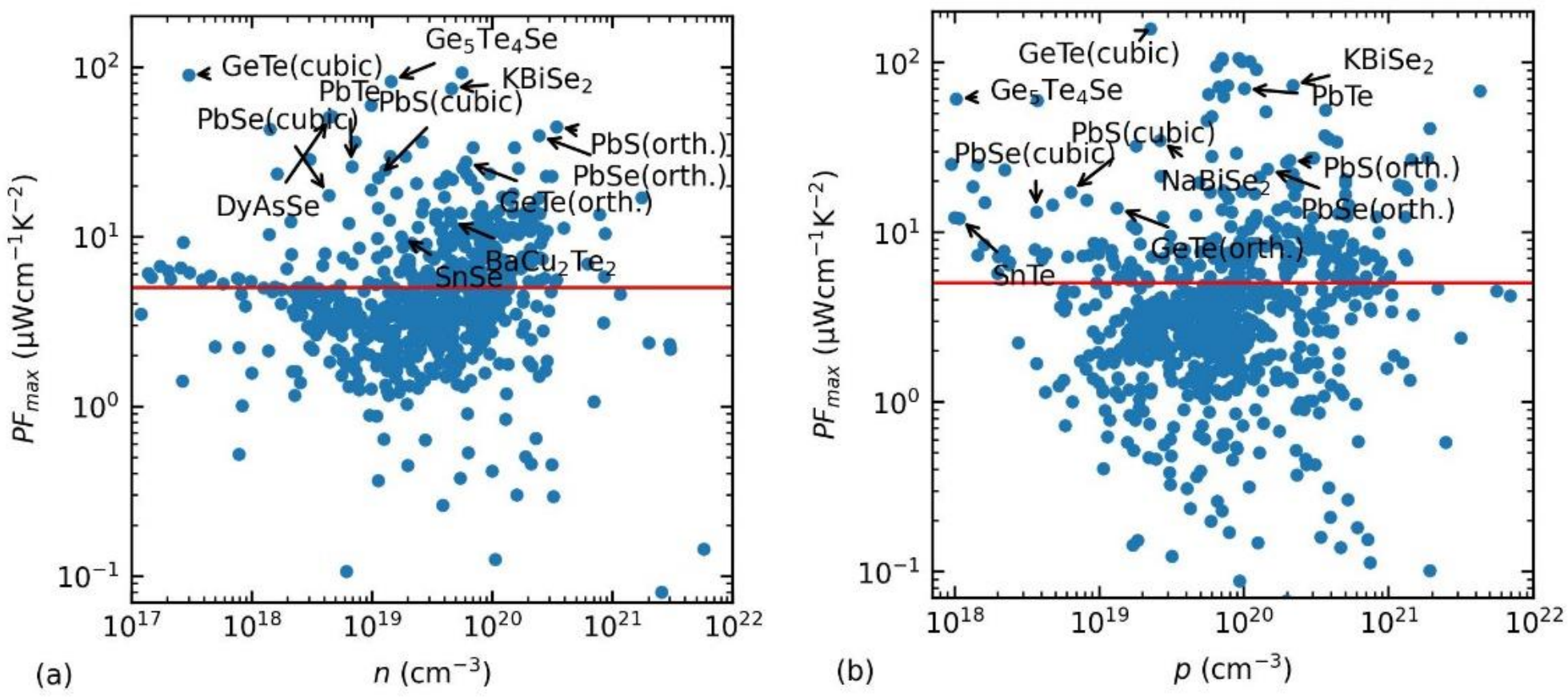

Figure 1. Maximum power factor as a function of the corresponding carrier concentration for the studied compounds in the temperature range from 300 K to 1000 K for (a) n type and (b) p type data. Some compounds with high power factor are marked.

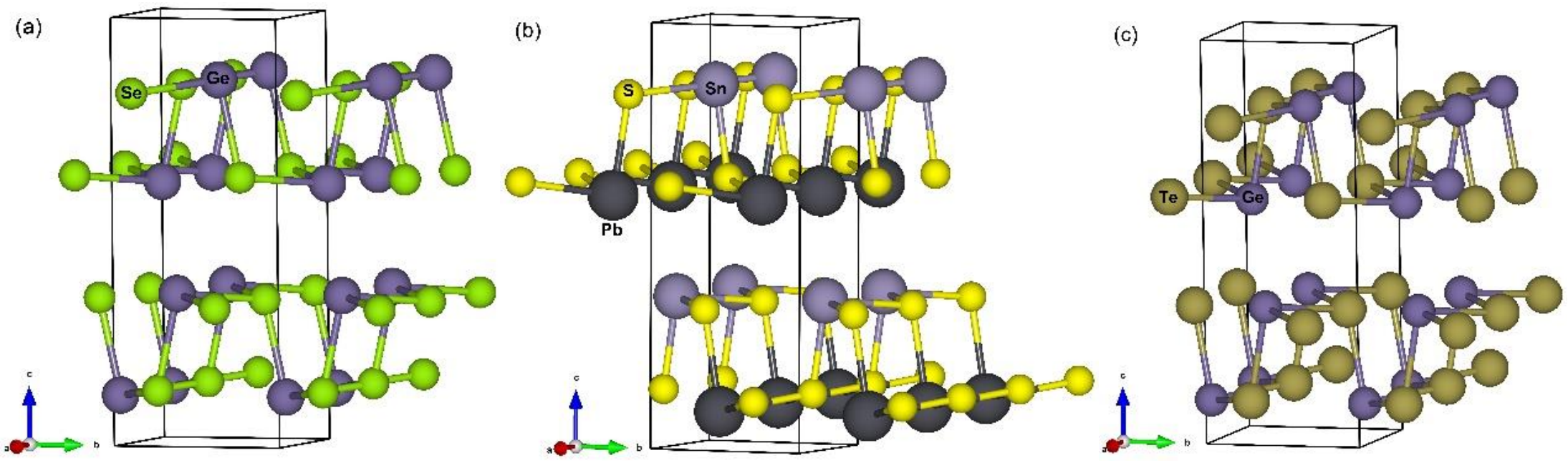

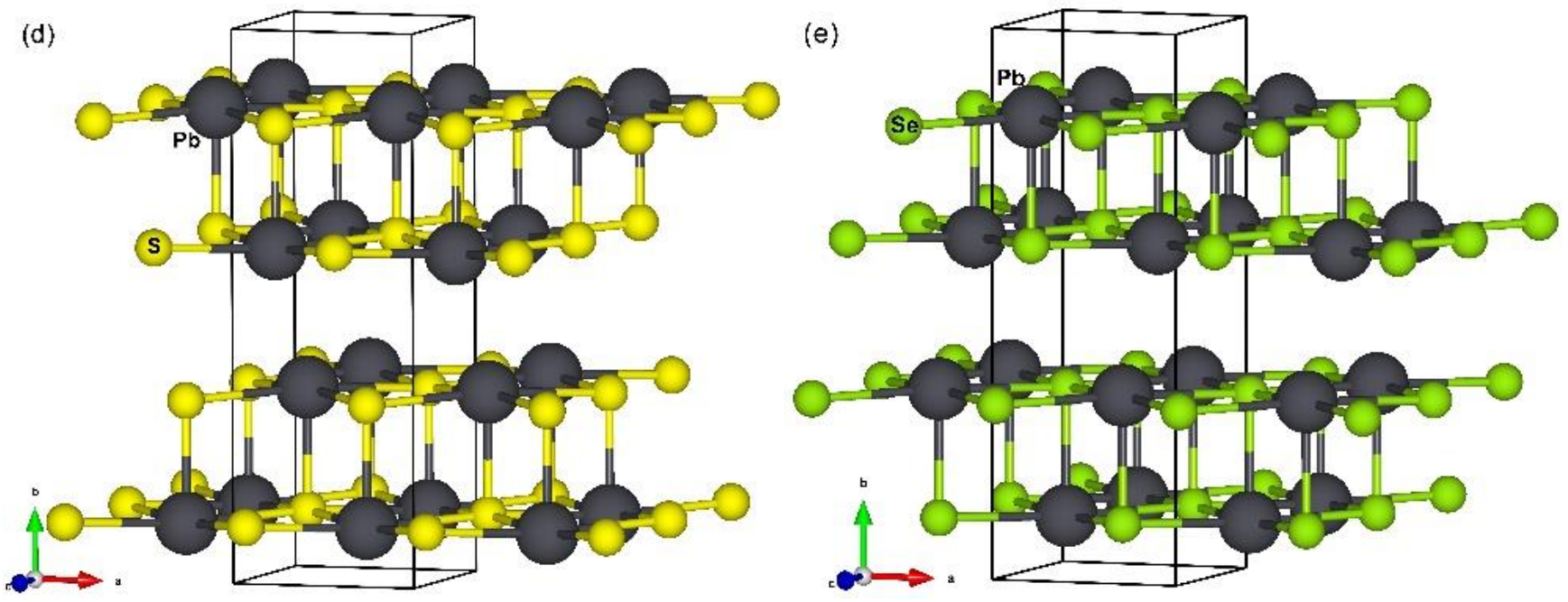

Figure 2. Crystal structures of (a) GeSe, (b) $SnPbS_2$, (c) GeTe, (d) PbS, (e) PbSe.

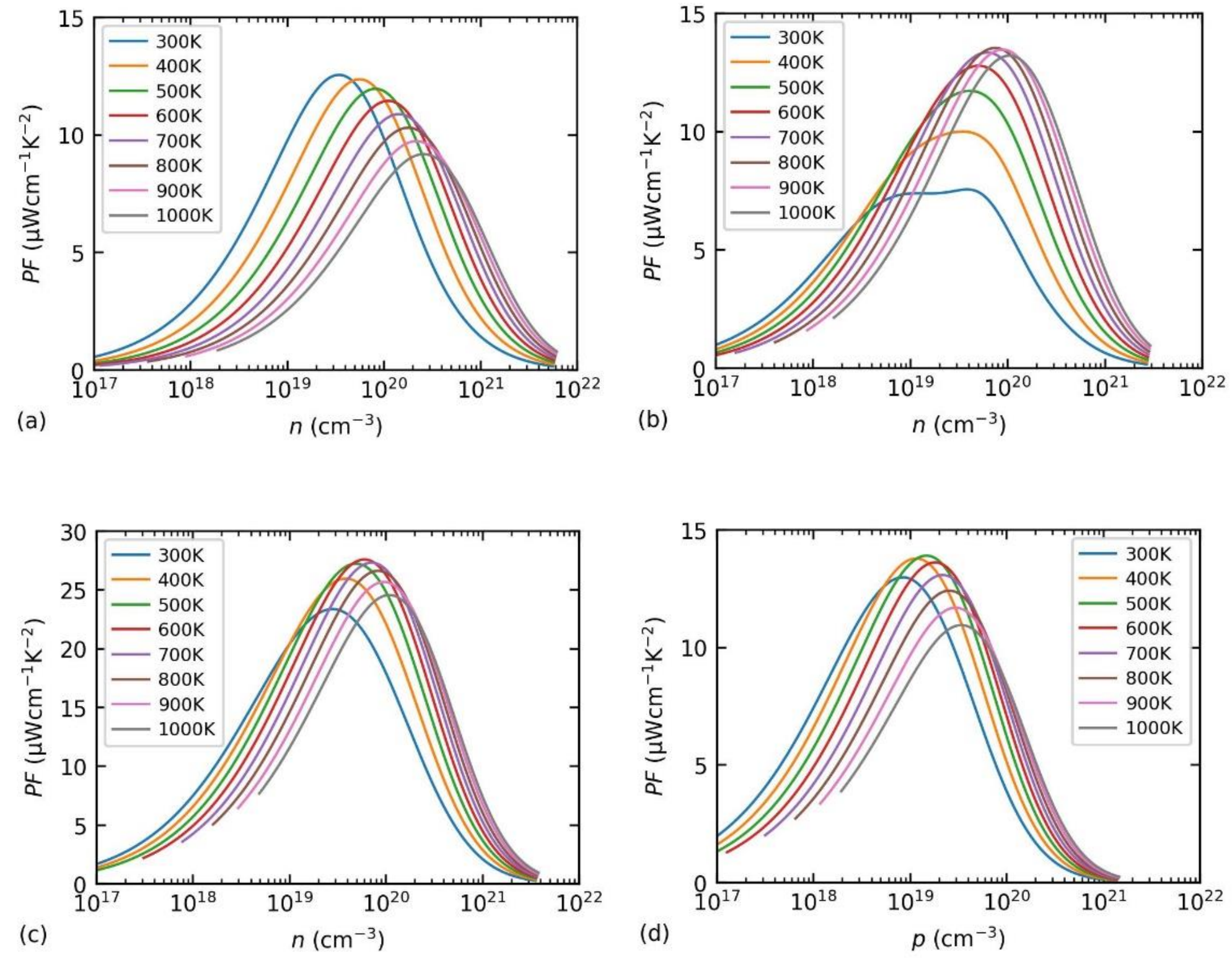

Figure 3. Power factor at varying temperatures and carrier concentrations for (a) GeSe, (b) $SnPbS_2$, and (c, d) GeTe.

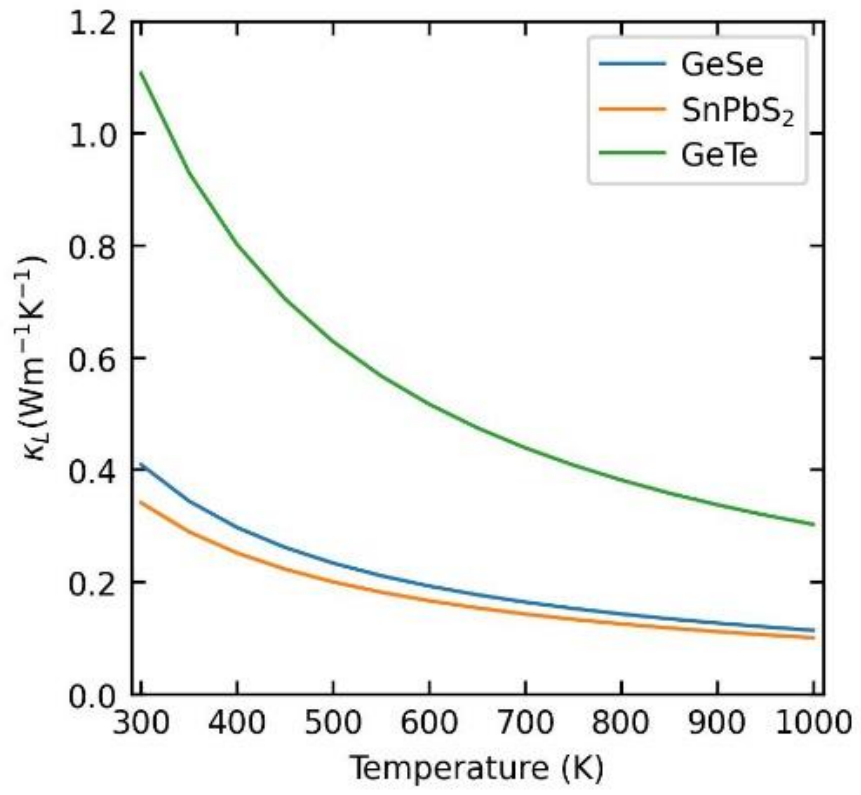

Figure 4. Lattice thermal conductivity of GeSe, $SnPbS_2$, and GeTe.

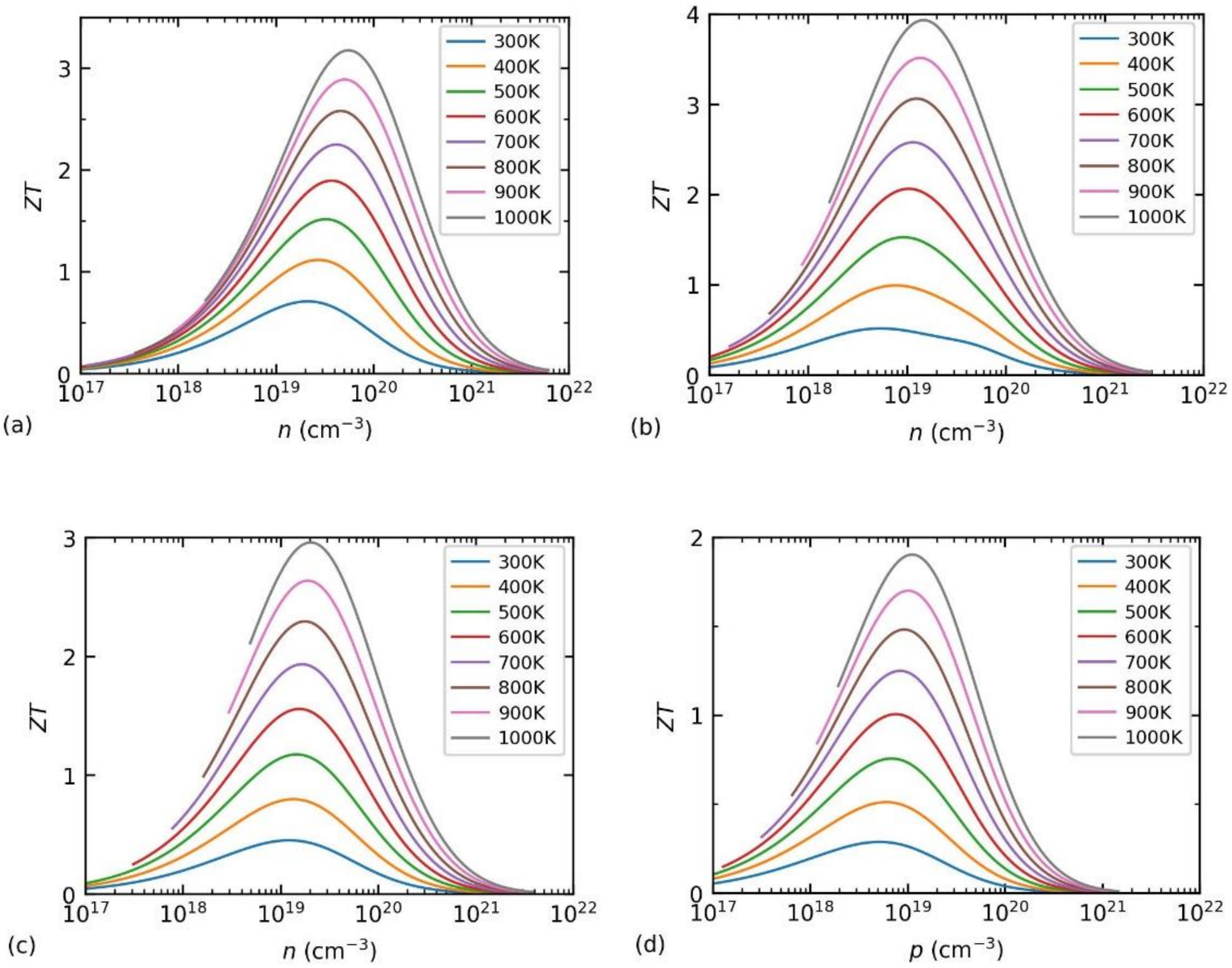

Figure 5. Figure of merit at varying temperatures and carrier concentrations for (a) GeSe, (b) $SnPbS_2$, and (c, d) GeTe.

**Ensemble learning models**

Ensemble learning combines the predictions of several base estimators built with a given learning algorithm in order to improve generalizability and robustness over a single estimator[61]. Multiple individual learners predict their own results and combined with some strategy such as weight average, voting. The performance of ensemble learning model is usually obviously better than a single best model. In this work, four common ensemble machine learning algorithms, including random forest (RF), gradient boosting decision trees (GBDT), adaptive boosting (AdaB) and extreme gradient boosting (XGB), combined with six kinds of input feature vectors (see detail in *Methods*), were used to train the classification models. Both the n type and the p type datasets were divided into 90% training and 10% test sets.

Table 2 shows the performance measure on test sets of the best model of each ensemble learning algorithm. The optimal hyperparameters for each model are listed in Table S4 in supplementary information. The complete table for each choice of input feature vector of each learning algorithm can also be found in supplementary information (Table S5 – S8). The accuracy of models trained on n type data are higher than 85%, while those trained on p type data are even higher than 90%. Since our objective is reducing the first-principles computational cost as much as possible, we prefer other metrics especially precision and recall. These two metrics directly reflect the efficiency of picking up truly good thermoelectric materials of a model. The models trained on n type data have precision higher than 80% (except the GBDT model), and recall higher than 90%. On the contrary, the models trained on p type data have precision higher than 90%, while recall higher than 75%. Therefore, F1, as the harmonic average of precision and recall, of these models are similar. From the column of AUC, all these models achieve AUC values larger than 0.9. Thus, all of them are very good classifiers. Generally, the performance metrics of different algorithms are similar for n type data and p type data, respectively. Although the best input feature vectors are different for different algorithms, GRDF_BOP appear four times in this table. It seems this kind feature vector is better than other used feature vectors generally.

Table 2. Performance measure on test sets of the best model of each ensemble learning algorithm with corresponding input feature vectors.

| | model | features | Acc. | Prec. | Recall | F1 | AUC |
|---|---|---|---|---|---|---|---|
| N | GBDT | GRDF_BOP | 0.86 | 0.78 | 0.90 | 0.84 | 0.97 |
| | XGB | VORONOI | 0.88 | 0.81 | 0.94 | 0.87 | 0.95 |
| | AdaB | PRDF16 | 0.89 | 0.85 | 0.90 | 0.88 | 0.95 |
| | RF | GRDF_BOP | 0.88 | 0.81 | 0.94 | 0.87 | 0.96 |
| P | GBDT | PRDF16 | 0.92 | 0.95 | 0.79 | 0.86 | 0.96 |
| | XGB | GRDF_BOP | 0.92 | 0.95 | 0.79 | 0.86 | 0.96 |
| | AdaB | PRDF25 | 0.89 | 0.90 | 0.75 | 0.82 | 0.93 |
| | RF | GRDF_BOP | 0.89 | 0.90 | 0.75 | 0.82 | 0.95 |

In the above trained models, the input feature vectors include band structure descriptors, such as band effective mass, band gap. Therefore, it is meaningful to compare the performance of the models with and without these descriptors in order to see if we can further reduce the computing cost. In supplementary information, Table S9 and S10 show the performance measure using GBDT and XGB algorithms together with input feature vectors with and without band structure parameters. Generally, compared with the performance of those models with input features having band structure parameters, the models without band structure parameters perform less well for almost all metrics we evaluated. Showing these band structure parameters are important for accurate predictions.

Machine learning is criticized for being a black box. Interpretability is always a hot topic in ML field. Here we used SHapley Additive exPlanations (SHAP) to explain our trained models[62]. SHAP connects optimal credit allocation with local explanations using the classic Shapley values from game theory and their related extensions. The most attracting point of SHAP is it can decompose the output of a model to the contribution of each input feature for each sample. Fig. 6 shows using SHAP to analyze the GBDT models presented in Table 2 (Fig. S6 in supplementary information shows the analysis of XGB models in Table 2). Fig. 6(a) and (b) show a summary plot of top ten important features — it takes the mean absolute SHAP value of each feature over all the samples of the dataset. All five band structure parameters are among top ten features for both n type and p type models, that explains why there is obvious difference between the performance of models with and without band structural parameters as input. Fig. 6(c) and (d) show beeswarm plotting to summarize the entire distribution of SHAP values for each feature, where the color of each point represents the feature value of that individual. These two pictures can reveal the direction of the feature's effect. For example, the higher band degeneracy $N_c$ or $N_v$ is, the higher SHAP value is, which means higher probability to be good thermoelectric materials. For conductivity effective mass, on the contrary, the lower $m_c^*$ is, the higher SHAP value is. This is consistent with our analysis in the last section. Another important feature revealed by SHAP analysis is deformation potential constant $\Xi$ — larger $\Xi$ leads to negative impact on SHAP value. This also matches with theory, since large $\Xi$ causes large electronic scattering rate, thus small carrier mobility. Therefore, our models truly capture the underline physics by just learning from a small amount of data.

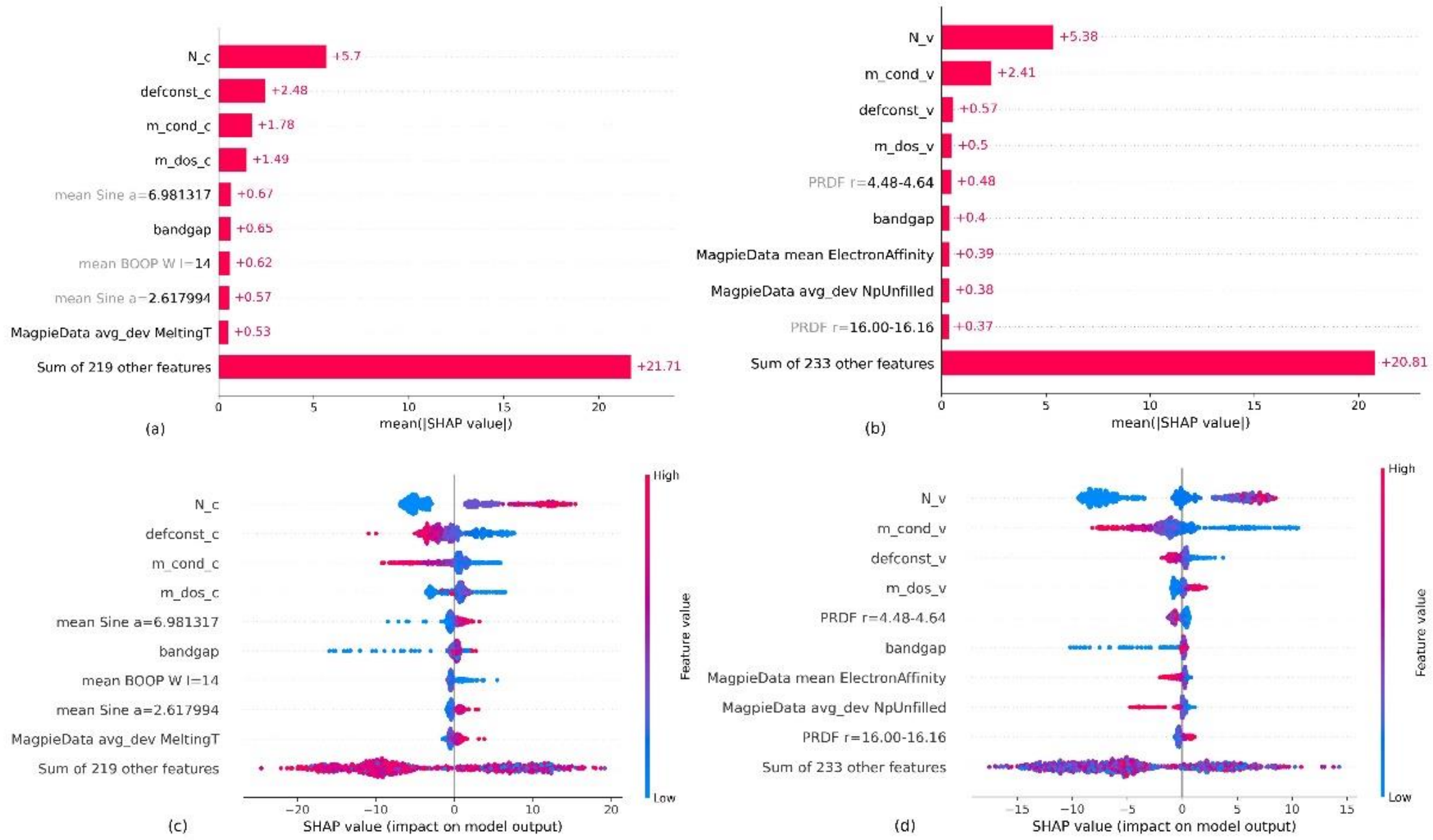

Figure 6. SHAP analysis based on GBDT models in Table 2. (a)(c) for n type model, (b)(d) for p type model. (a)(b) bar chart of the average SHAP value magnitude for each input feature. (c)(d) a set of beeswarm plots, where each dot corresponds to a sample. The dot's position on the $x$ axis shows the impact that feature has on the model's prediction for that sample. When multiple dots locate at the same $x$ position, they pile up to show density.

**Deep learning models**

Deep learning models are mostly used in computer vision and nature language processing fields, since these fields have accumulated a large amount of data. Recently, as the development of several large materials database, such as Materials Project, OQMD and AFLOW, deep learning models for materials science come forth in large numbers. Different from ensemble learning, one of the advantages of deep learning model is it learns the representation of materials by itself. Graph neural network is one of the mostly used deep learning models in the materials science. Graphs are a powerful non-Euclidean data structure method for establishing relationships between nodes and their edges. Thus, it is natural to represent a crystal structure as a graph. In this work, we used both MEGNet and M3GNet to predict the label of the candidate materials. Compared with MEGNet, M3GNet also incorporates three-body interaction by building an additional line graph for bonds, which makes it one of the state-of-the-art models[63].

Instead training the model from the beginning, it is better to make use of those pretrained models and do transfer learning. Both MEGNet and M3GNet have pretrained models which were trained on larger dataset from MP to predict formation energy. Here we reused their element embedding layers since such embedding should be universal for all kinds of tasks. After the readout stage, the node features for each atom were combined into the crystal feature vector. At this time, we added into the band structure descriptors, and the concatenated vector was pass to a multi-layer perceptron (MLP) to predict the target value. Both the n type and the p type datasets were divided into 80% training, 10% validation and 10% test sets. The hyperparameters, including learning rate, number of units in each layer, weight decay etc., were optimized based on the loss of the validation set. Then, the final models were trained on the datasets merged by the training sets and validation sets. Table S11 and S12 list the settings for hyperparameters. Fig. S7 shows the training loss and test loss during the final training process. The test loss converged at the end of training for all selected models.

Beside training deep models directly, we also used the pretrained M3GNet model as a way to generate structure feature vectors, then combined with composition features and band structure features as input to GBDT and XGB models. By doing this, we want to see if features learned from one task can be applicable to another.

The performance measure on test sets for two deep models and two models combined ensemble learning algorithm and M3GNet structure feature vectors are shown in Table 3. Although the datasets are small, the performance of the two deep models are pretty good. Especially, the M3GNet model for n type data has accuracy, precision and recall all higher than 90%, which is the best model we've obtained. For p type data, although M3GNet model is not much better than those in Table 2, its performance is more balanced — both precision and recall higher than 80%. The performance of MEGNet models for both n type and p type data are a bit less than M3GNet models. In addition, models trained on n type data are generally better than those trained on p type data, mainly because the p type dataset is more unbalanced than the n type dataset. For the two combinatorial models, their performance are not better than deep models. Moreover, compared with GBDT and XGB models with general purpose structure features (see Table S5 and S6), these two models don't show clear advantages, suggesting the trained structure features for one task may not be the optimal choice for another. Finally, we combined four ensemble learning models in Table 2 with M3GNet model in Table 3 to form a voting classifier. In this classifier, the predicted class label for a particular sample is the class label that represents the majority of the class labels predicted by each individual classifier. Although the voting classifiers are not better than the best individual classifier for both n type and p type data, they should be more robust for the unseen data.

In figure 7, the ROC curves which calculated on n type and p type test sets of four ensemble learning models in Table 2 and two deep learning models in Table 3 are shown. The AUC values of all these models are higher than 0.9, closing to the perfect classifier.

Table 3. Performance measure on test sets of the deep learning models and combinatorial models with deep learned input structure features.

| | model | Acc. | Prec. | Recall | F1 | AUC |
|---|---|---|---|---|---|---|
| N | MEGNet | 0.88 | 0.82 | 0.90 | 0.86 | 0.96 |
| | M3GNet | 0.93 | 0.91 | 0.94 | 0.92 | 0.96 |
| | GBDT-M3G | 0.80 | 0.71 | 0.87 | 0.78 | 0.95 |
| | XGB-M3G | 0.84 | 0.74 | 0.94 | 0.83 | 0.94 |
| | Voting | 0.87 | 0.80 | 0.90 | 0.85 | |
| P | MEGNet | 0.87 | 0.77 | 0.83 | 0.80 | 0.91 |
| | M3GNet | 0.88 | 0.80 | 0.83 | 0.82 | 0.92 |
| | GBDT-M3G | 0.83 | 0.72 | 0.75 | 0.73 | 0.93 |
| | XGB-M3G | 0.86 | 0.76 | 0.79 | 0.78 | 0.90 |
| | Voting | 0.92 | 0.95 | 0.79 | 0.86 | |

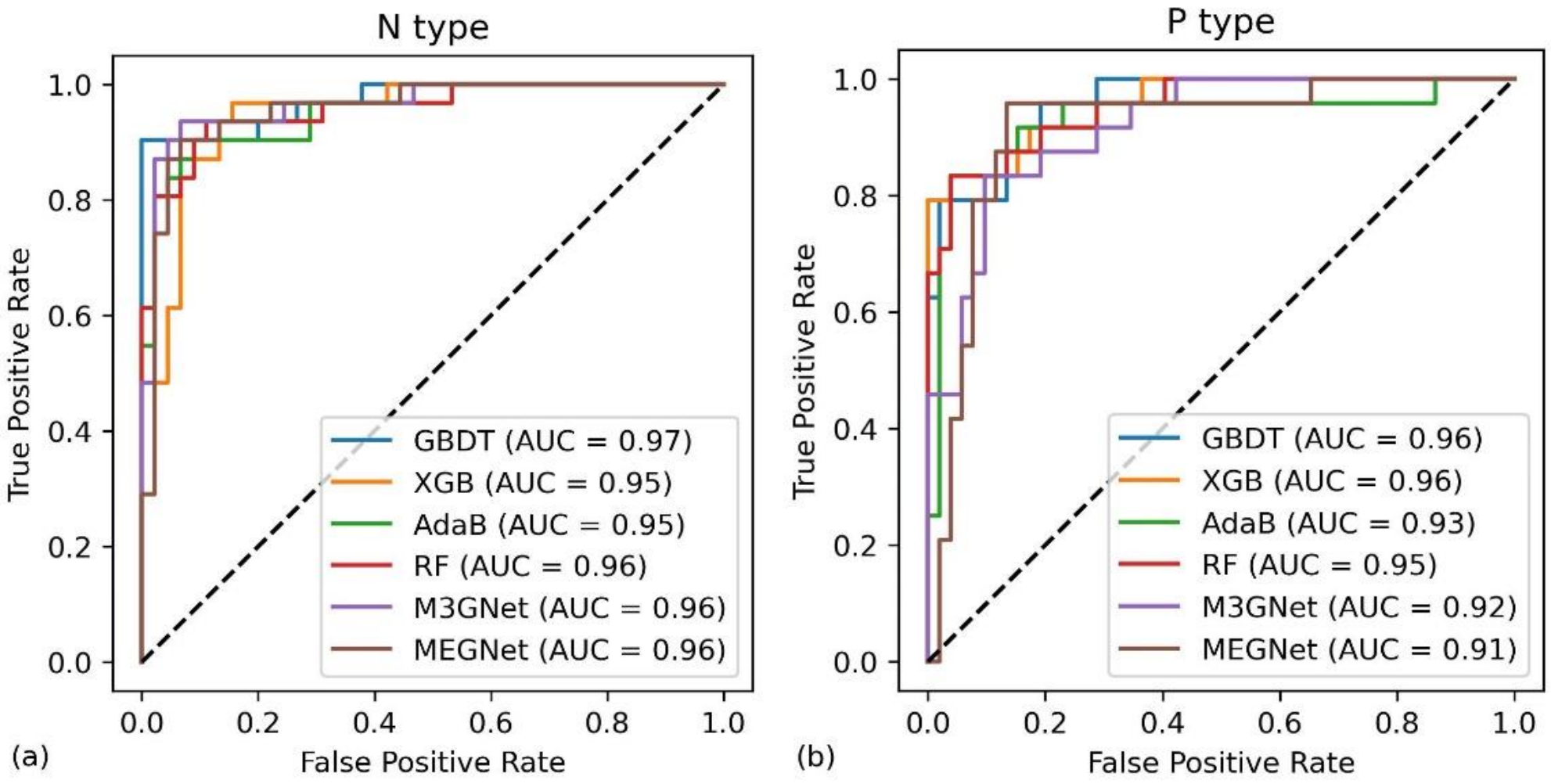

Figure 7. Receiver Operating Characteristic (ROC) curves of models listed in Table 2 and Table 3. The black dash line is the chance level line (AUC = 0.5).

## CONCLUSIONS

In this work, we did first-principles transport properties calculations for a large number of compounds extracted from public database. A database containing 796 compounds' thermoelectric related transport properties were built, and it includes many novel and promising TE materials we have found. Some of them, such as $Ge_5Te_4Se$, $KBiSe_2$, GeTe (*Pnma*) and $BaCu_2Te_2$, could have their performance better than those currently state-of-the-art TE materials. Then, in order to furtherly accelerate the high-throughput search of novel TE materials, we trained several types of ML models to classify the candidate compounds for n type and p type doping, respectively. First, four ensemble learning algorithms combined with six kinds of input feature vectors were trained and compared for n type and p type data, respectively. The results on test sets show that ensemble learning algorithms with proper input feature vectors can achieve classification accuracy higher than 85% and the precision and the recall could be higher than 90%. The latter two metrics are more important since we are more interested in picking up positive samples. Moreover, two graph neural network models were also trained for the same tasks. Although deep learning models are usually used with big data, with optimized hyperparameters our trained deep models' performance are the same good as those ensemble learning models on the test datasets. Especially, the M3GNet model for n type data achieve accuracy, precision and recall all higher than 90%, which is the best among the models we obtained. All of the trained models achieve the AUC higher than 0.9, and the ROC curves of them are close to that of perfect classifier. In order to understand the trained models, the SHAP method was used to analyze those ensemble models. The Interpretability analysis show that these models truly capture the underline physics to some extent. For example, they correctly reflect the direction of the band degeneracy and band effective mass' effects. All of the presented models can be used together to do prescreening of thermoelectric materials, which improves the generalizability and robustness for the unseen data. Finally, combining ML prediction and high-throughput first-principles calculation together will greatly accelerate the process of discovering promising thermoelectric materials.

## DATA AVAILABILITY

The datasets used to generate the ML models can be found online at https://github.com/Baijianlu/ML4TE.git.

## CODE AVAILABILITY

The codes and trained models in this work can be found online at https://github.com/Baijianlu/ML4TE.git.

## ACKNOWLEDGEMENTS

We acknowledge the usage of the Skoltech HPC cluster ARKUDA for obtaining the results presented in this paper.

## COMPETING INTERESTS

The authors declare no competing interests.

## REFERENCES

[1] Rowe, D. M. (Ed.). (2018). *Thermoelectrics handbook: macro to nano*. CRC press.

[2] Tan, G., Zhao, L. D., & Kanatzidis, M. G. (2016). Rationally designing high-performance bulk thermoelectric materials. *Chemical reviews*, *116*(19), 12123-12149.

[3] Shi, X., Chen, L., & Uher, C. (2016). Recent advances in high-performance bulk thermoelectric materials. *International Materials Reviews*, *61*(6), 379-415.

[4] Bell, L. E. (2008). Cooling, heating, generating power, and recovering waste heat with thermoelectric systems. *Science*, *321*(5895), 1457-1461.

[5] Jaziri, N., Boughamoura, A., Müller, J., Mezghani, B., Tounsi, F., & Ismail, M. (2020). A comprehensive review of Thermoelectric Generators: Technologies and common applications. *Energy Reports*, *6*, 264–287.

[6] Nozariasbmarz, A., Collins, H., Dsouza, K., Polash, M. H., Hosseini, M., Hyland, M., ... & Vashaee, D. (2020). Review of wearable thermoelectric energy harvesting: From body temperature

to electronic systems. *Applied Energy*, *258*, 114069.
[7] Zhao, L. D., Hao, S., Lo, S. H., Wu, C. I., Zhou, X., Lee, Y., ... & Kanatzidis, M. G. (2013). High thermoelectric performance via hierarchical compositionally alloyed nanostructures. *Journal of the American Chemical Society*, *135*(19), 7364-7370.
[8] Zhu, T., Liu, Y., Fu, C., Heremans, J. P., Snyder, J. G., & Zhao, X. (2017). Compromise and synergy in high-efficiency thermoelectric materials. *Advanced materials*, *29*(14), 1605884.
[9] Gorai, P., Stevanović, V., & Toberer, E. S. (2017). Computationally guided discovery of thermoelectric materials. *Nature Reviews Materials*, *2*(9), 1-16.
[10] Hao, S., Dravid, V. P., Kanatzidis, M. G., & Wolverton, C. (2019). Computational strategies for design and discovery of nanostructured thermoelectrics. *npj Computational Materials*, *5*(1), 58.
[11] Gutiérrez Moreno, J. J., Cao, J., Fronzi, M., & Assadi, M. H. N. (2020). A review of recent progress in thermoelectric materials through computational methods. *Materials for Renewable and Sustainable Energy*, *9*(3), 16.
[12] Zhou, J. J., Park, J., Lu, I. T., Maliyov, I., Tong, X., & Bernardi, M. (2021). Perturbo: A software package for ab initio electron–phonon interactions, charge transport and ultrafast dynamics. *Computer Physics Communications*, 264, 107970.
[13] Askarpour, V., & Maassen, J. (2023). First-principles analysis of intravalley and intervalley electron-phonon scattering in thermoelectric materials. *Physical Review B*, *107*(4), 045203.
[14] Li, W., Carrete, J., Katcho, N. A., & Mingo, N. (2014). ShengBTE: A solver of the Boltzmann transport equation for phonons. *Computer Physics Communications*, *185*(6), 1747-1758.
[15] Pal, K., Xia, Y., Shen, J., He, J., Luo, Y., Kanatzidis, M. G., & Wolverton, C. (2021). Accelerated discovery of a large family of quaternary chalcogenides with very low lattice thermal conductivity. *npj Computational Materials*, *7*(1), 82.
[16] Zhu, H., Hautier, G., Aydemir, U., Gibbs, Z. M., Li, G., Bajaj, S., ... & Ceder, G. (2015). Computational and experimental investigation of TmAgTe2 and XYZ2 compounds, a new group of thermoelectric materials identified by first-principles high-throughput screening. *Journal of Materials Chemistry C*, *3*(40), 10554–10565.
[17] Carrete, J., Mingo, N., Wang, S., & Curtarolo, S. (2014). Nanograined Half-Heusler Semiconductors as Advanced Thermoelectrics: An Ab Initio High-Throughput Statistical Study. *Advanced Functional Materials*, *24*(47), 7427-7432.
[18] Xi, L., Pan, S., Li, X., Xu, Y., Ni, J., Sun, X., ... & Zhang, W. (2018). Discovery of high-performance thermoelectric chalcogenides through reliable high-throughput material screening. *Journal of the American Chemical Society*, *140*(34), 10785–10793.
[19] Gorai, P., Parilla, P., Toberer, E. S., & Stevanovic, V. (2015). Computational exploration of the binary A1B1 chemical space for thermoelectric performance. *Chemistry of Materials*, *27*(18), 6213–6221.
[20] Jia, T., Feng, Z., Guo, S., Zhang, X., & Zhang, Y. (2020). Screening promising thermoelectric materials in binary chalcogenides through high-throughput computations. *ACS applied materials & interfaces*, *12*(10), 11852–11864.
[21] LeCun, Y., Bengio, Y., & Hinton, G. (2015). Deep learning. *nature*, *521*(7553), 436-444.
[22] Karniadakis, G. E., Kevrekidis, I. G., Lu, L., Perdikaris, P., Wang, S., & Yang, L. (2021). Physics-informed machine learning. *Nature Reviews Physics*, *3*(6), 422-440.
[23] Choudhary, K., DeCost, B., Chen, C., Jain, A., Tavazza, F., Cohn, R., ... & Wolverton, C. (2022). Recent advances and applications of deep learning methods in materials science. *npj Computational*

*Materials*, *8*(1), 59.
[24] Reiser, P., Neubert, M., Eberhard, A., Torresi, L., Zhou, C., Shao, C., ... & Friederich, P. (2022). Graph neural networks for materials science and chemistry. *Communications Materials*, *3*(1), 93.
[25] Wang, H., Fu, T., Du, Y., Gao, W., Huang, K., Liu, Z., ... & Zitnik, M. (2023). Scientific discovery in the age of artificial intelligence. *Nature*, *620*(7972), 47-60.
[26] Ryan, K., Lengyel, J., & Shatruk, M. (2018). Crystal structure prediction via deep learning. *Journal of the American Chemical Society*, *140*(32), 10158-10168.
[27] Podryabinkin, E. V., Tikhonov, E. V., Shapeev, A. V., & Oganov, A. R. (2019). Accelerating crystal structure prediction by machine-learning interatomic potentials with active learning. *Physical Review B*, *99*(6), 064114.
[28] Behler, J. (2021). Four generations of high-dimensional neural network potentials. *Chemical Reviews*, *121*(16), 10037-10072.
[29] Unke, O. T., Chmiela, S., Sauceda, H. E., Gastegger, M., Poltavsky, I., Schütt, K. T., ... & Müller, K. R. (2021). Machine learning force fields. *Chemical Reviews*, *121*(16), 10142-10186.
[30] Zubatiuk, T., & Isayev, O. (2021). Development of multimodal machine learning potentials: Toward a physics-aware artificial intelligence. *Accounts of Chemical Research*, *54*(7), 1575-1585.
[31] Lee, J., Seko, A., Shitara, K., Nakayama, K., & Tanaka, I. (2016). Prediction model of band gap for inorganic compounds by combination of density functional theory calculations and machine learning techniques. *Physical Review B*, *93*(11), 115104.
[32] Bartók, A. P., De, S., Poelking, C., Bernstein, N., Kermode, J. R., Csányi, G., & Ceriotti, M. (2017). Machine learning unifies the modeling of materials and molecules. *Science advances*, *3*(12), e1701816.
[33] Li, M., Dai, L., & Hu, Y. (2022). Machine learning for harnessing thermal energy: From materials discovery to system optimization. *ACS energy letters*, *7*(10), 3204-3226.
[34] Gan, Y., Wang, G., Zhou, J., & Sun, Z. (2021). Prediction of thermoelectric performance for layered IV-V-VI semiconductors by high-throughput ab initio calculations and machine learning. *npj Computational Materials*, *7*(1), 176.
[35] Lee, Y. L., Lee, H., Kim, T., Byun, S., Lee, Y. K., Jang, S., ... & Im, J. (2022). Data-driven enhancement of ZT in SnSe-based thermoelectric systems. *Journal of the American Chemical Society*, *144*(30), 13748-13763.
[36] Li, Y., Zhang, J., Zhang, K., Zhao, M., Hu, K., & Lin, X. (2022). Large Data Set-Driven Machine Learning Models for Accurate Prediction of the Thermoelectric Figure of Merit. *ACS Applied Materials & Interfaces*, *14*(50), 55517-55527.
[37] Jia, X., Yao, H., Yang, Z., Shi, J., Yu, J., Shi, R., ... & Liu, X. (2023). Advancing thermoelectric materials discovery through semi-supervised learning and high-throughput calculations. *Applied Physics Letters*, *123*(20).
[38] Ren, Q., Chen, D., Rao, L., Lun, Y., Tang, G., & Hong, J. (2024). Machine-learning-assisted discovery of 212-Zintl-phase compounds with ultra-low lattice thermal conductivity. *Journal of Materials Chemistry A*, *12*(2), 1157-1165.
[39] Luo, Y., Li, M., Yuan, H., Liu, H., & Fang, Y. (2023). Predicting lattice thermal conductivity via machine learning: A mini review. *npj Computational Materials*, *9*(1), 4.
[40] Jia, X., Deng, Y., Bao, X., Yao, H., Li, S., Li, Z., ... & Liu, X. (2022). Unsupervised machine learning for discovery of promising half-Heusler thermoelectric materials. *npj Computational Materials*, *8*(1), 34.

[41] Fan, T., & Oganov, A. R. (2021). AICON2: A program for calculating transport properties quickly and accurately. *Computer Physics Communications*, *266*, 108027.
[42] Chen, T., & Guestrin, C. (2016, August). Xgboost: A scalable tree boosting system. In Proceedings of the 22nd acm sigkdd international conference on knowledge discovery and data mining (pp. 785-794).
[43] Chen, C., Ye, W., Zuo, Y., Zheng, C., & Ong, S. P. (2019). Graph networks as a universal machine learning framework for molecules and crystals. *Chemistry of Materials*, *31*(9), 3564-3572.
[44] Chen, C., & Ong, S. P. (2022). A universal graph deep learning interatomic potential for the periodic table. *Nature Computational Science*, *2*(11), 718-728.
[45] Kresse, G., & Furthmüller, J. (1996). Efficient iterative schemes for ab initio total-energy calculations using a plane-wave basis set. *Physical review B*, *54*(16), 11169.
[46] Kresse, G., & Furthmüller, J. (1996). Efficiency of ab-initio total energy calculations for metals and semiconductors using a plane-wave basis set. *Computational materials science*, *6*(1), 15–50.
[47] Perdew, J. P., Burke, K., & Ernzerhof, M. (1996). Generalized gradient approximation made simple. *Physical review letters*, *77*(18), 3865.
[48] Gonze, X., & Lee, C. (1997). Dynamical matrices, Born effective charges, dielectric permittivity tensors, and interatomic force constants from density-functional perturbation theory. *Physical Review B*, *55*(16), 10355.
[49] Setyawan, W., & Curtarolo, S. (2010). High-throughput electronic band structure calculations: Challenges and tools. *Computational materials science*, *49*(2), 299-312.
[50] Ong, S. P., Richards, W. D., Jain, A., Hautier, G., Kocher, M., Cholia, S., ... & Ceder, G. (2013). Python Materials Genomics (pymatgen): A robust, open-source python library for materials analysis. *Computational Materials Science*, *68*, 314–319.
[51] Mathew, K., Montoya, J. H., Faghaninia, A., Dwarakanath, S., Aykol, M., Tang, H., ... & Jain, A. (2017). Atomate: A high-level interface to generate, execute, and analyze computational materials science workflows. *Computational Materials Science*, *139*, 140–152.
[52] Jain, A., Ong, S. P., Chen, W., Medasani, B., Qu, X., Kocher, M., ... & Persson, K. A. (2015). FireWorks: A dynamic workflow system designed for high-throughput applications. *Concurrency and Computation: Practice and Experience*, *27*(17), 5037–5059.
[53] Fan, T., & Oganov, A. R. (2021). Discovery of high performance thermoelectric chalcogenides through first-principles high-throughput screening. *Journal of Materials Chemistry C*, *9*(38), 13226-13235.
[54] Ward, L., Agrawal, A., Choudhary, A., & Wolverton, C. (2016). A general-purpose machine learning framework for predicting properties of inorganic materials. *npj Computational Materials*, *2*(1), 1-7.
[55] Ward, L., Liu, R., Krishna, A., Hegde, V. I., Agrawal, A., Choudhary, A., & Wolverton, C. (2017). Including crystal structure attributes in machine learning models of formation energies via Voronoi tessellations. *Physical Review B*, *96*(2), 024104.
[56] Schütt, K. T., Glawe, H., Brockherde, F., Sanna, A., Müller, K. R., & Gross, E. K. (2014). How to represent crystal structures for machine learning: Towards fast prediction of electronic properties. *Physical Review B*, *89*(20), 205118.
[57] Seko, A., Hayashi, H., Nakayama, K., Takahashi, A., & Tanaka, I. (2017). Representation of compounds for machine-learning prediction of physical properties. *Physical Review B*, *95*(14), 144110.

[58] Ward, L., Dunn, A., Faghaninia, A., Zimmermann, N. E., Bajaj, S., Wang, Q., ... & Jain, A. (2018). Matminer: An open source toolkit for materials data mining. *Computational Materials Science*, *152*, 60-69.
[59] Pedregosa, F., Varoquaux, G., Gramfort, A., Michel, V., Thirion, B., Grisel, O., ... & Duchesnay, É. (2011). Scikit-learn: Machine learning in Python. *the Journal of machine Learning research*, *12*, 2825-2830.
[60] Jain, A., Ong, S. P., Hautier, G., Chen, W., Richards, W. D., Dacek, S., ... & Persson, K. A. (2013). Commentary: The Materials Project: A materials genome approach to accelerating materials innovation. *APL materials*, *1*(1).
[61] Zhou, Z. H. (2021). *Machine learning*. Springer nature.
[62] Lundberg, S. M., Erion, G., Chen, H., DeGrave, A., Prutkin, J. M., Nair, B., ... & Lee, S. I. (2020). From local explanations to global understanding with explainable AI for trees. *Nature machine intelligence*, *2*(1), 56-67.
[63] Riebesell, J., Goodall, R. E., Jain, A., Benner, P., Persson, K. A., & Lee, A. A. (2023). Matbench Discovery--An evaluation framework for machine learning crystal stability prediction. *arXiv preprint arXiv:2308.14920*.
[64] Zhao, L. D., Lo, S. H., Zhang, Y., Sun, H., Tan, G., Uher, C., ... & Kanatzidis, M. G. (2014). Ultralow thermal conductivity and high thermoelectric figure of merit in SnSe crystals. *nature*, *508*(7496), 373-377.

Supplementary Information

**Combining Machine Learning Models with First-Principles High-Throughput Calculation to Accelerate the Search of Promising Thermoelectric Materials**

Tao Fan*, Artem R. Oganov
Skolkovo Institute of Science and Technology, Bolshoy Boulevard 30, bld. 1, 121205 Moscow, Russia.

**Labeling the dataset**

Since our aim is to train a classification model to distinguish good and not good thermoelectric materials, the first step is labeling each sample in n type and p type datasets, respectively. Currently we have the maximum power factor ($PF_{max}$) value in the temperature range 300 K to 1000 K, which can represent the capability of electronic transport properties of a compound. In order to label the sample, we need to decide the boundary value of $PF_{max}$ which separates the samples as good (label 1) and not good (label 0). Recall the definition of figure of merit $ZT$,

$$ZT = \frac{\alpha^2 \sigma T}{\kappa} = \frac{PF \cdot T}{\kappa}$$

thus

$$PF = \frac{ZT \cdot \kappa}{T}$$

Usually, a compound with $ZT$ larger than 1 is thought as good thermoelectric material. The thermal conductivity $\kappa$ include lattice part ($\kappa_L$) and electron part ($\kappa_e$). Here we assume only lattice thermal conductivity matters and omit the contribution from the electrons. The lowest lattice thermal conductivity a compound can reach is so called amorphous limit, which is possible when the phonons have a mean free path $l$ on the order of the interatomic spacing[1,2]. The value of the amorphous limit is usually between 0.2 $W \cdot m^{-1} \cdot K^{-1}$ and 0.5 $W \cdot m^{-1} \cdot K^{-1}$. Here we use the lower boundary 0.2 $W \cdot m^{-1} \cdot K^{-1}$. Then, we can calculate the lowest possible $PF_{max}$ which could lead to $ZT$ larger than 1 in the temperature range of 300 K to 1000 K,

$$\frac{1 \cdot 0.2}{1000} = 0.0002 \leq PF_{max} \leq \frac{1 \cdot 0.2}{300} = 0.00067 \Leftrightarrow$$
$$2\ \mu W \cdot cm^{-1} \cdot K^{-2} \leq PF_{max} \leq 6.7\ \mu W \cdot cm^{-1} \cdot K^{-2}$$

Choosing which value exactly as the boundary for good and not good thermoelectric materials is heuristic. If selecting a value close to 2 $\mu W \cdot cm^{-1} \cdot K^{-2}$, we actually lower the standard and expand the searching space, but at the risk of including many compounds that are actually not promising and wasting our resources. If selecting a value close to 6.7 $\mu W \cdot cm^{-1} \cdot K^{-2}$, we lift the standard and narrow the search space, but also at the risk of missing some promising compounds. Here we use the boundary value as 5 $\mu W \cdot m^{-1} \cdot K^{-2}$, because, practically it is rare to get such a low kappa value as 0.2 $W \cdot m^{-1} \cdot K^{-1}$, and also we are more interested in materials showing good performance near room temperature. According to this boundary value, for n type dataset, #positive:#negative

= 308:444, while for p type dataset, #positive:#negative = 244:513.

Table S1. Elemental properties used to compute composition descriptors

| | | | | |
|---|---|---|---|---|
| Atomic Number | Mendeleev Number | Atomic Weight | Melting Temperature | Column |
| Row | Covalent Radius | Electron Negativity | # s Valence Electrons | # p Valence Electrons |
| # d Valence Electrons | # f Valence Electrons | Total # Valance Electrons | # Unfilled s States | # Unfilled p States |
| # Unfilled d States | # Unfilled f States | Total # Unfilled States | Specific Volume of 0 K Ground State | Band Gap Energy of 0 K Ground State |
| Electron Affinity | First Ionization Energy | Space Group Number of 0 K Ground State | | |

Table S2. Top 50 non-cubic n type thermoelectric materials found in this work. Entry id refers to the id number in Materials Project database. $N$ is the band degeneracy, $m_c^*$ the conductivity effective mass, $m_d^*$ the density of states effective mass, $E_g$ the band gap, and $\Xi$ the deformation potential constant. $PF_{max}$ was calculated within temperature range from 300 K to 1000 K

| Formula | Entry id | $N$ | $m_c^*$ | $m_d^*$ | $E_g$ (eV) | $\Xi$ (eV) | $PF_{max}$ ($\mu W \cdot cm^{-1} \cdot K^{-2}$) |
|---|---|---|---|---|---|---|---|
| $Ge_5Te_4Se$ | mp-1224356 | 12 | 0.129 | 0.533 | 0.226 | 11.022 | 82.032 |
| $KBiSe_2$ | mp-36539 | 8 | 0.226 | 1.485 | 0.814 | 7.098 | 74.726 |
| TbAsSe | mp-1102476 | 2 | 0.046 | 0.104 | 0.125 | 4.166 | 59.381 |
| DyAsSe | mp-1102952 | 2 | 0.031 | 0.073 | 0.073 | 4.826 | 50.953 |
| YAsSe | mp-1095603 | 2 | 0.037 | 0.087 | 0.096 | 5.167 | 50.362 |
| PbS | mp-1018115 | 4 | 0.187 | 0.320 | 0.915 | 8.562 | 44.409 |
| HoAsSe | mp-1212091 | 2 | 0.024 | 0.052 | 0.032 | 4.710 | 43.052 |
| PbSe | mp-1063670 | 4 | 0.186 | 0.339 | 0.578 | 8.079 | 39.249 |
| TbAsS | mp-1101828 | 3 | 0.120 | 0.185 | 0.054 | 9.167 | 35.943 |
| $YSe_2$ | mp-1232213 | 2 | 0.362 | 1.323 | 0.145 | 3.000 | 33.333 |
| SmAsSe | mp-1208883 | 2 | 0.060 | 0.166 | 0.133 | 9.067 | 28.309 |
| GeTe | mp-1080459 | 4 | 0.151 | 0.260 | 0.454 | 7.901 | 27.554 |
| $Ga_2Te_5$ | mp-2371 | 6 | 0.211 | 0.478 | 0.944 | 8.554 | 25.120 |
| YAsS | mp-1102959 | 2 | 0.045 | 0.374 | 0.056 | 10.117 | 24.806 |
| $ErAsSe_4$ | mp-1213048 | 2 | 0.023 | 0.048 | 0.024 | 7.196 | 23.454 |
| $Te_4Pb_5Se$ | mp-1217406 | 4 | 0.172 | 0.479 | 0.590 | 10.084 | 22.717 |
| $Bi_2Se_2S$ | mp-1227504 | 4 | 0.375 | 0.719 | 0.721 | 8.396 | 22.620 |
| $TlGaTe_2$ | mp-3785 | 6 | 0.225 | 0.455 | 0.532 | 8.761 | 21.611 |
| $Tl_9SbTe_6$ | mp-34292 | 5 | 0.198 | 0.730 | 0.634 | 8.077 | 20.452 |
| $NdTl_2InTe_4$ | mp-1220095 | 10 | 0.276 | 1.027 | 0.968 | 8.991 | 19.229 |
| $RbDy_2Ag_3Te_5$ | mp-1190502 | 2 | 0.343 | 0.466 | 1.007 | 3.706 | 17.909 |
| $Tl_9SbSe_6$ | mp-676274 | 5 | 0.305 | 0.990 | 0.728 | 7.443 | 17.111 |
| $BaTe_2$ | mp-2150 | 4 | 0.345 | 1.275 | 0.343 | 7.306 | 17.080 |
| $Ba_2HfS_4$ | mp-9321 | 1 | 0.264 | 0.904 | 0.872 | 4.502 | 16.683 |
| $Tl_9BiSe_6$ | mp-34361 | 5 | 0.287 | 0.844 | 0.688 | 7.403 | 16.403 |
| $TlCu_7S_4$ | mp-21964 | 6 | 0.557 | 1.398 | 0.641 | 9.973 | 15.062 |
| $BiSbSe_3$ | mp-1227508 | 4 | 0.305 | 0.555 | 0.638 | 8.233 | 14.871 |
| ErPS | mp-1191596 | 1 | 0.306 | 0.491 | 0.1137 | 4.485 | 14.762 |
| $LuCuPbSe_3$ | mp-1205378 | 2 | 0.382 | 0.619 | 0.996 | 8.820 | 14.169 |
| $Dy_3CuSe_6$ | mp-1225646 | 1 | 0.107 | 0.307 | 0.660 | 8.321 | 13.846 |
| $Ba_2ZrS_4$ | mp-3813 | 1 | 0.264 | 1.002 | 0.631 | 5.067 | 13.785 |
| $Pr_2InCuS_5$ | mp-1220176 | 4 | 0.639 | 1.661 | 0.941 | 10.550 | 13.756 |
| $Sm_3CuSe_6$ | mp-1219332 | 1 | 0.121 | 0.337 | 0.733 | 7.700 | 13.733 |
| $Ba_5Hf_4S_{13}$ | mp-557032 | 1 | 0.261 | 0.677 | 0.664 | 5.142 | 13.672 |
| $SnPbS_2$ | mp-1218951 | 4 | 0.216 | 0.361 | 0.626 | 8.475 | 13.518 |
| GeS | mp-12910 | 2 | 0.318 | 1.352 | 0.602 | 7.472 | 13.429 |
| PtS | mp-288 | 5 | 0.464 | 0.901 | 0.407 | 14.723 | 13.411 |
| $Sr_3Sb_4S_9$ | mp-29295 | 2 | 0.381 | 1.574 | 1.014 | 6.183 | 13.409 |
| $LiCuS_2$ | mp-766486 | 4 | 0.612 | 1.159 | 0.546 | 7.738 | 12.989 |
| $CuSbSe_2$ | mp-20331 | 4 | 0.396 | 0.668 | 0.510 | 9.099 | 12.819 |
| $Bi_2Se_3$ | mp-23164 | 4 | 0.256 | 0.527 | 0.607 | 8.475 | 12.794 |
| GeSe | mp-700 | 2 | 0.258 | 1.237 | 0.867 | 7.181 | 12.542 |
| $Ba_4Hf_3S_{10}$ | mp-1189858 | 1 | 0.258 | 0.669 | 0.704 | 5.176 | 12.535 |
| TlSe | mp-1836 | 4 | 0.212 | 0.543 | 0.197 | 7.317 | 12.486 |
| $BaCu_2Te_2$ | mp-30133 | 4 | 0.384 | 0.766 | 0.602 | 8.109 | 12.420 |
| $LuCuPbSe_3$ | mp-583052 | 3 | 0.434 | 0.603 | 1.053 | 7.586 | 12.389 |
| $ZnCu_2SiTe_4$ | mp-1078498 | 3 | 0.133 | 0.210 | 0.166 | 12.644 | 12.218 |
| NaTe | mp-28353 | 2 | 0.196 | 0.373 | 0.520 | 6.282 | 12.026 |
| $Er_2Te_3$ | mp-14643 | 2 | 0.306 | 0.376 | 0.601 | 5.392 | 11.793 |
| $TmCuPbSe_3$ | mp-865269 | 2 | 0.406 | 0.662 | 1.041 | 8.600 | 11.421 |

Table S3. Top 50 non-cubic p type thermoelectric materials found in this work. Entry id refers to the id number in Materials Project database. $N$ is the band degeneracy, $m_c^*$ the conductivity effective mass, $m_d^*$ the density of states effective mass, $E_g$ the band gap, and $\Xi$ the deformation potential constant. $PF_{max}$ was calculated within temperature range from 300 K to 1000 K

| Formula | Entry id | $N$ | $m_c^*$ | $m_d^*$ | $E_g$ (eV) | $\Xi$ (eV) | $PF_{max}$ ($\mu W \cdot cm^{-1} \cdot K^{-2}$) |
|---|---|---|---|---|---|---|---|
| $KBiSe_2$ | mp-36539 | 12 | 0.356 | 2.423 | 0.814 | 7.264 | 73.372 |
| $Ge_5Te_4Se$ | mp-1224356 | 4 | 0.034 | 0.112 | 0.226 | 13.397 | 61.018 |
| TbAsS | mp-1101828 | 2 | 0.034 | 0.110 | 0.054 | 5.117 | 59.956 |
| $NaBiSe_2$ | mp-35015 | 4 | 0.151 | 0.649 | 0.679 | 9.070 | 34.868 |
| $Ba_2HfS_4$ | mp-9321 | 6 | 0.878 | 2.902 | 0.872 | 5.903 | 34.083 |
| $NaSbS_2$ | mp-1173513 | 4 | 0.053 | 0.417 | 0.089 | 11.117 | 32.156 |
| $LiBiS_2$ | mp-33526 | 4 | 0.159 | 0.517 | 0.811 | 10.942 | 27.934 |
| $Ba_2ZrS_4$ | mp-3813 | 6 | 0.888 | 2.887 | 0.631 | 6.346 | 27.435 |
| $Ba_4Hf_3S_{10}$ | mp-1189858 | 6 | 0.776 | 2.182 | 0.704 | 7.148 | 27.416 |
| PbS | mp-1018115 | 2 | 0.211 | 0.535 | 0.915 | 10.235 | 26.501 |
| DyAsSe | mp-1102952 | 2 | 0.036 | 0.073 | 0.073 | 9.236 | 25.257 |
| YAsSe | mp-1095603 | 2 | 0.043 | 0.089 | 0.096 | 9.417 | 24.958 |
| PbSe | mp-1063670 | 4 | 0.156 | 0.385 | 0.579 | 16.364 | 23.621 |
| TbAsSe | mp-1102476 | 2 | 0.057 | 0.110 | 0.125 | 9.367 | 23.303 |
| $Ba_4Zr_3S_{10}$ | mp-14883 | 6 | 0.787 | 2.152 | 0.485 | 7.780 | 22.199 |
| $Te_4Pb_5Se$ | mp-1217406 | 4 | 0.134 | 0.443 | 0.590 | 12.333 | 21.496 |
| $Sm_{10}Se_{19}$ | mp-29832 | 4 | 0.398 | 2.587 | 0.269 | 9.367 | 21.099 |
| $Ba_3Zr_2S_7$ | mp-9179 | 6 | 0.809 | 2.443 | 0.528 | 8.359 | 19.701 |
| $Er_2Se_3$ | mp-1225508 | 4 | 0.517 | 1.317 | 0.230 | 7.780 | 19.586 |
| $Ba_3Zr_2S_7$ | mp-554172 | 6 | 0.809 | 2.443 | 0.528 | 8.359 | 19.396 |
| $MgS_2$ | mp-1185953 | 6 | 0.822 | 3.948 | 0.932 | 8.628 | 19.136 |
| HoAsSe | mp-1212091 | 2 | 0.026 | 0.052 | 0.032 | 9.311 | 18.522 |
| DyPS | mp-1192185 | 3 | 0.613 | 1.015 | 0.295 | 7.270 | 17.628 |
| TbPS | mp-1190952 | 3 | 0.657 | 1.082 | 0.355 | 7.150 | 17.436 |
| GeS | mp-12910 | 4 | 0.137 | 0.307 | 0.602 | 10.118 | 17.146 |
| YPS | mp-1191026 | 3 | 0.634 | 1.061 | 0.307 | 7.450 | 17.118 |
| CoAsS | mp-4627 | 4 | 0.814 | 1.612 | 0.942 | 13.671 | 16.943 |
| $Zn_4CdSe_5$ | mp-1215624 | 5 | 0.803 | 1.757 | 0.920 | 7.913 | 16.667 |
| SnS | mp-559676 | 4 | 0.146 | 0.344 | 0.365 | 12.498 | 15.443 |
| ErAsSe | mp-1213048 | 2 | 0.025 | -0.048 | 0.024 | 9.111 | 14.926 |
| $Tm_{10}S_{19}$ | mp-1204382 | 4 | 0.968 | 6.831 | 0.552 | 9.850 | 14.523 |
| | | | | | | | |
| $SnTe_4Pb_3$ | mp-1218925 | 4 | 0.051 | 0.146 | 0.060 | 11.683 | 14.484 |
| ErTeAs | mp-1212634 | 3 | 0.262 | 2.011 | 0.248 | 10.310 | 14.180 |
| GeTe | mp-1080459 | 2 | 0.086 | 0.215 | 0.454 | 11.641 | 13.893 |
| SmPS | mp-1191565 | 3 | 1.077 | 1.775 | 0.380 | 6.267 | 13.608 |
| $Li_5SbS$ | mp-767409 | 2 | 0.284 | 0.857 | 0.605 | 7.331 | 13.561 |
| DyTeAs | mp-1212803 | 3 | 0.265 | 2.188 | 0.225 | 10.271 | 13.204 |
| $BaAg_2GeS_4$ | mp-7394 | 5 | 0.753 | 1.842 | 0.585 | 11.884 | 12.880 |
| $YbY_2Se_4$ | mp-1193988 | 2 | 0.485 | 2.337 | 1.122 | 7.667 | 12.633 |
| $Ga_2HgTe_4$ | mp-1224839 | 3 | 0.406 | 1.534 | 0.326 | 9.780 | 12.507 |
| $SnPb_4S_5$ | mp-1218954 | 4 | 0.156 | 0.649 | 0.085 | 10.568 | 12.341 |
| $K_2Se_3$ | mp-7670 | 6 | 1.536 | 8.316 | 0.627 | 4.302 | 12.318 |
| $GeTe_2Pb$ | mp-1224318 | 2 | 0.031 | 0.055 | 0.067 | 10.468 | 12.205 |
| $YbTm_2Se_4$ | mp-1193091 | 2 | 0.459 | 1.244 | 1.140 | 7.700 | 11.607 |
| $YbEr_2Se_4$ | mp-1192445 | 2 | 0.462 | 1.248 | 1.129 | 7.667 | 11.389 |
| $YbHo_2Se_4$ | mp-1194224 | 2 | 0.464 | 1.251 | 1.117 | 7.633 | 11.221 |
| $CaIn_2Te_4$ | mp-677072 | 5 | 0.425 | 1.310 | 0.653 | 10.178 | 10.823 |
| NdPS | mp-1191667 | 3 | 1.562 | 2.775 | 0.310 | 6.144 | 10.738 |
| SmTeAs | mp-1208842 | 3 | 0.265 | 1.219 | 0.167 | 10.056 | 10.475 |
| $Sm_2S_3$ | mp-1403 | 3 | 0.780 | 1.330 | 0.750 | 8.350 | 10.438 |

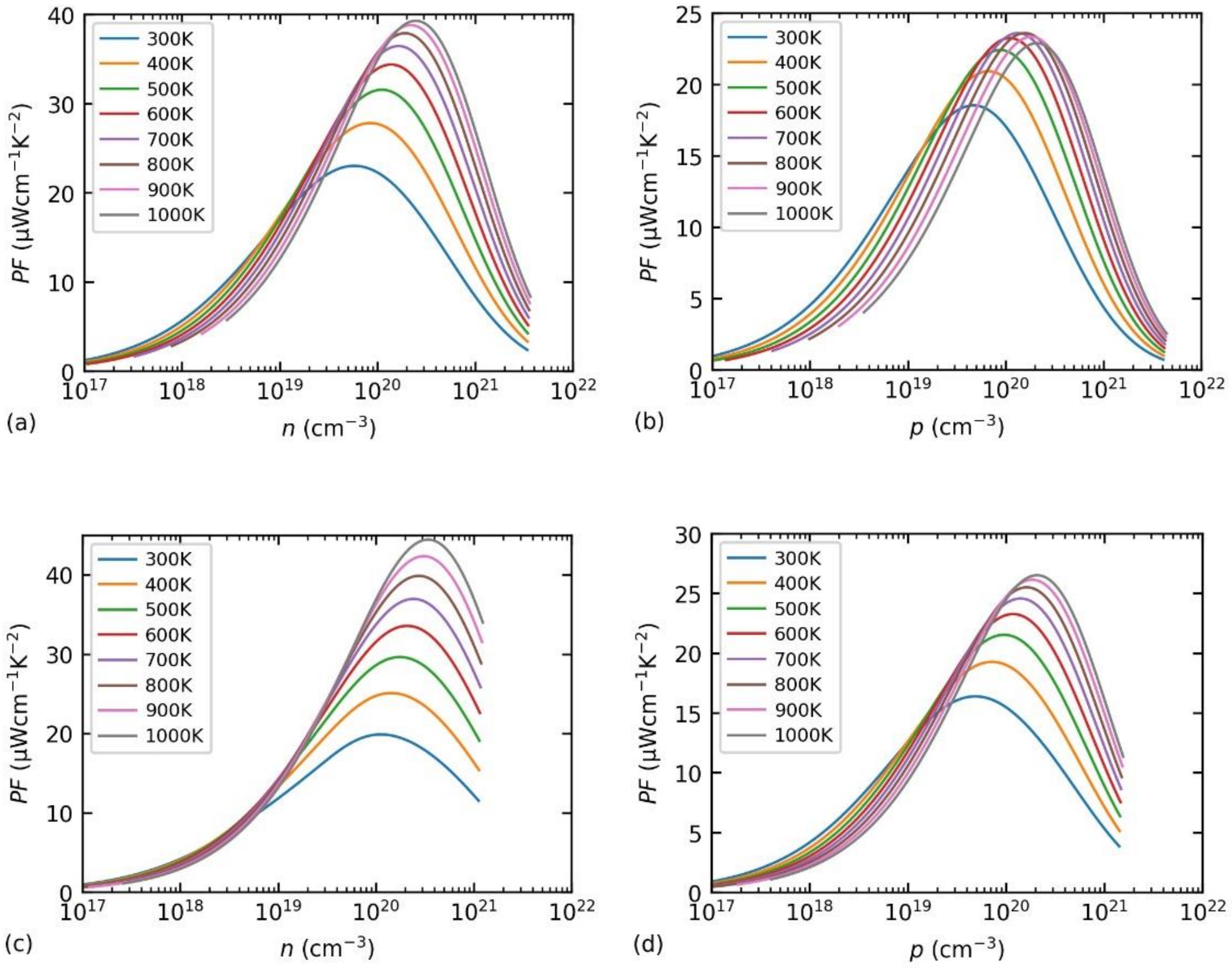

Figure S1. Power factor at varying temperatures and carrier concentrations for (a, b) PbSe and (c, d) PbS.

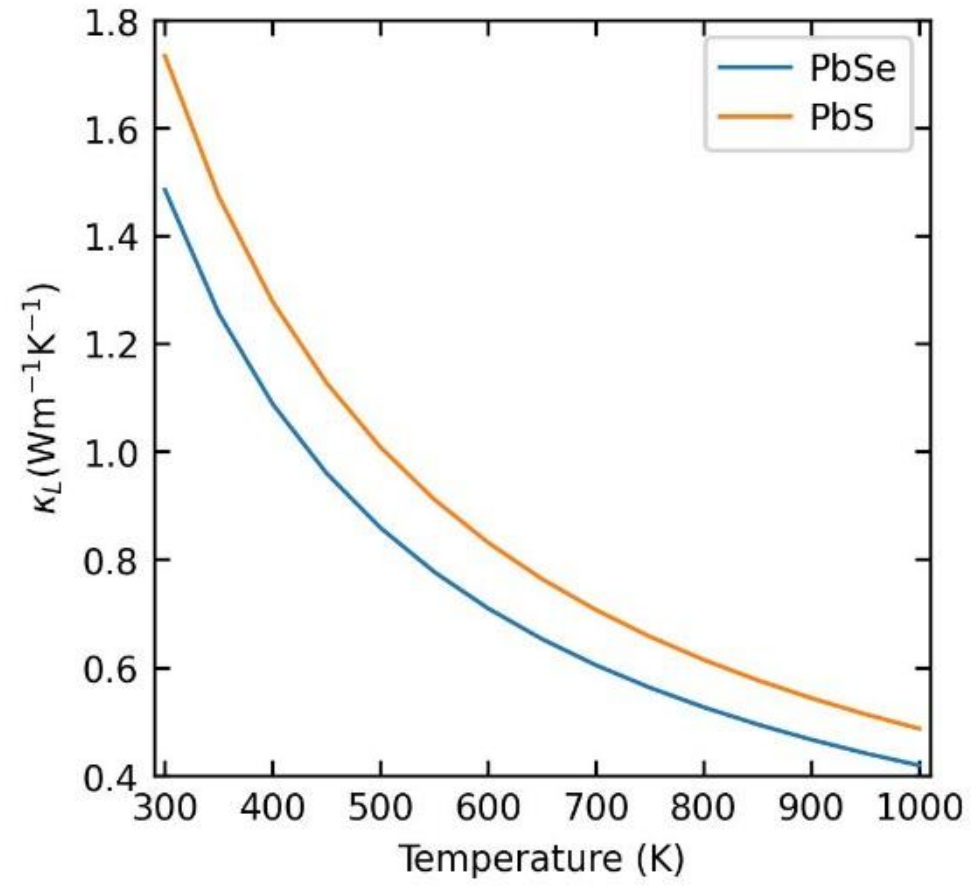

Figure S2. Lattice thermal conductivity of PbSe and PbS.

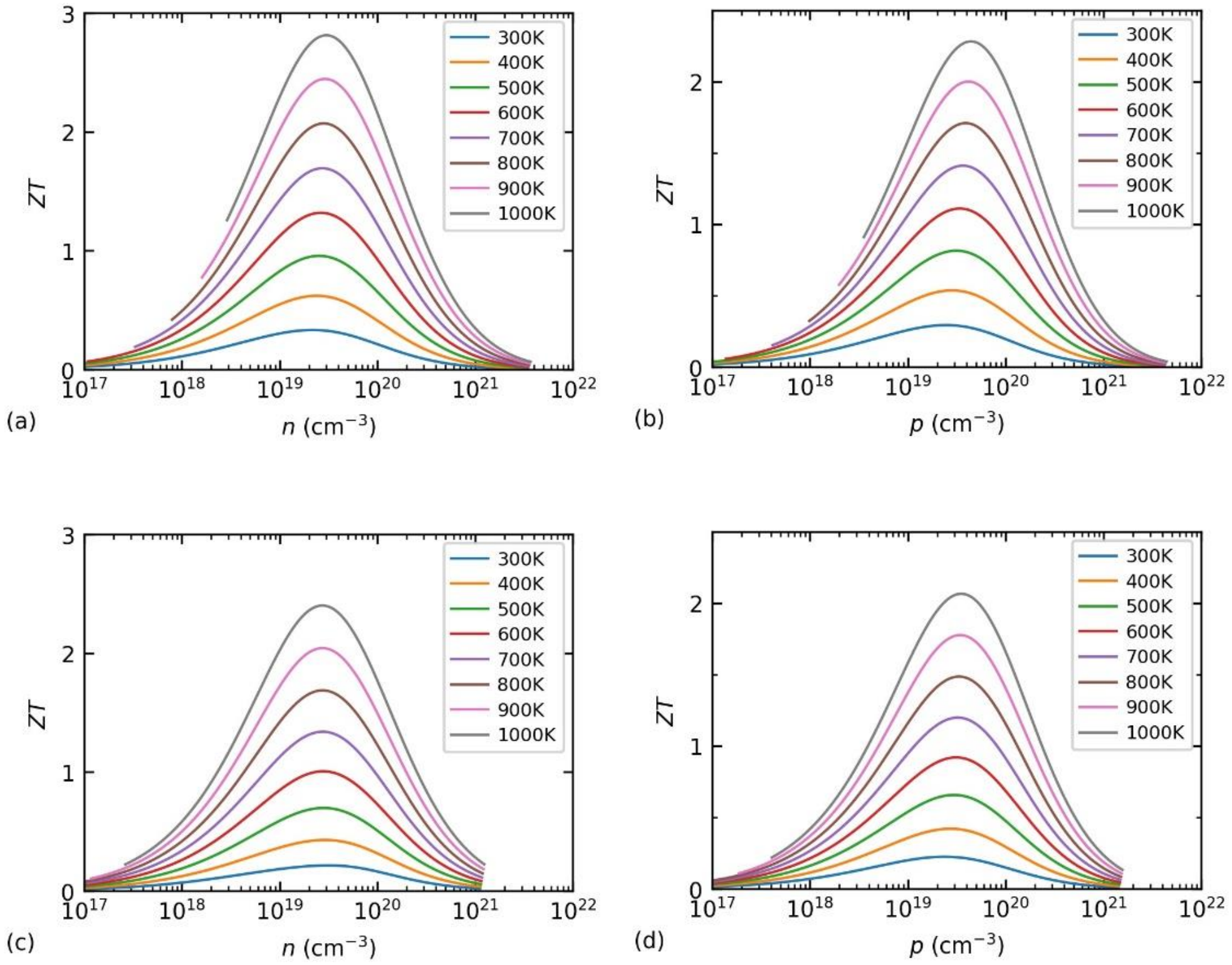

Figure S3. Figure of merit at varying temperatures and carrier concentrations for (a, b) PbSe and (c, d) PbS.

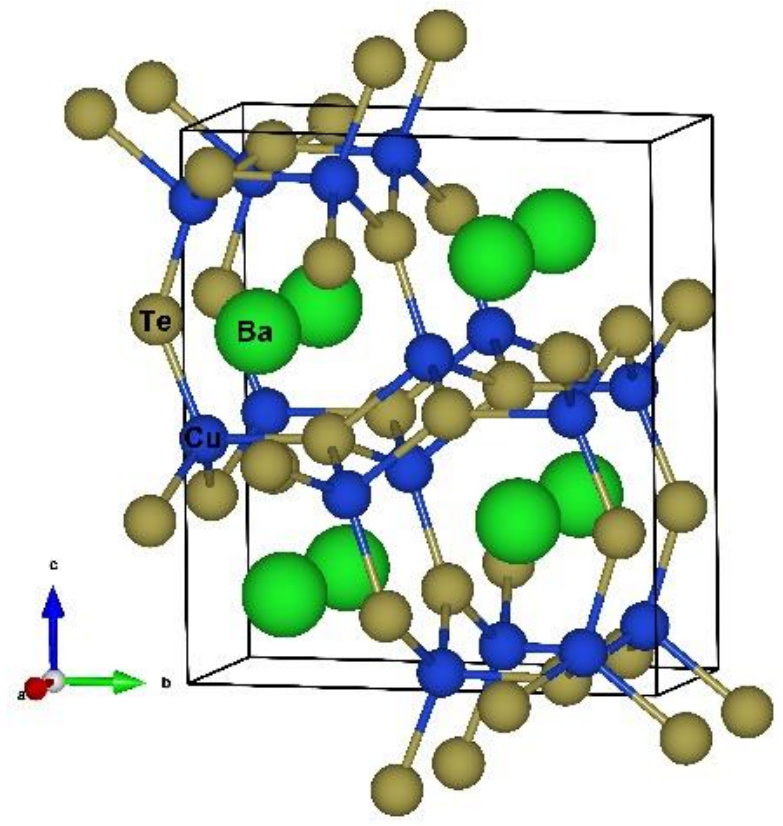

Figure S4. Crystal structures of $BaCu_2Te_2$.

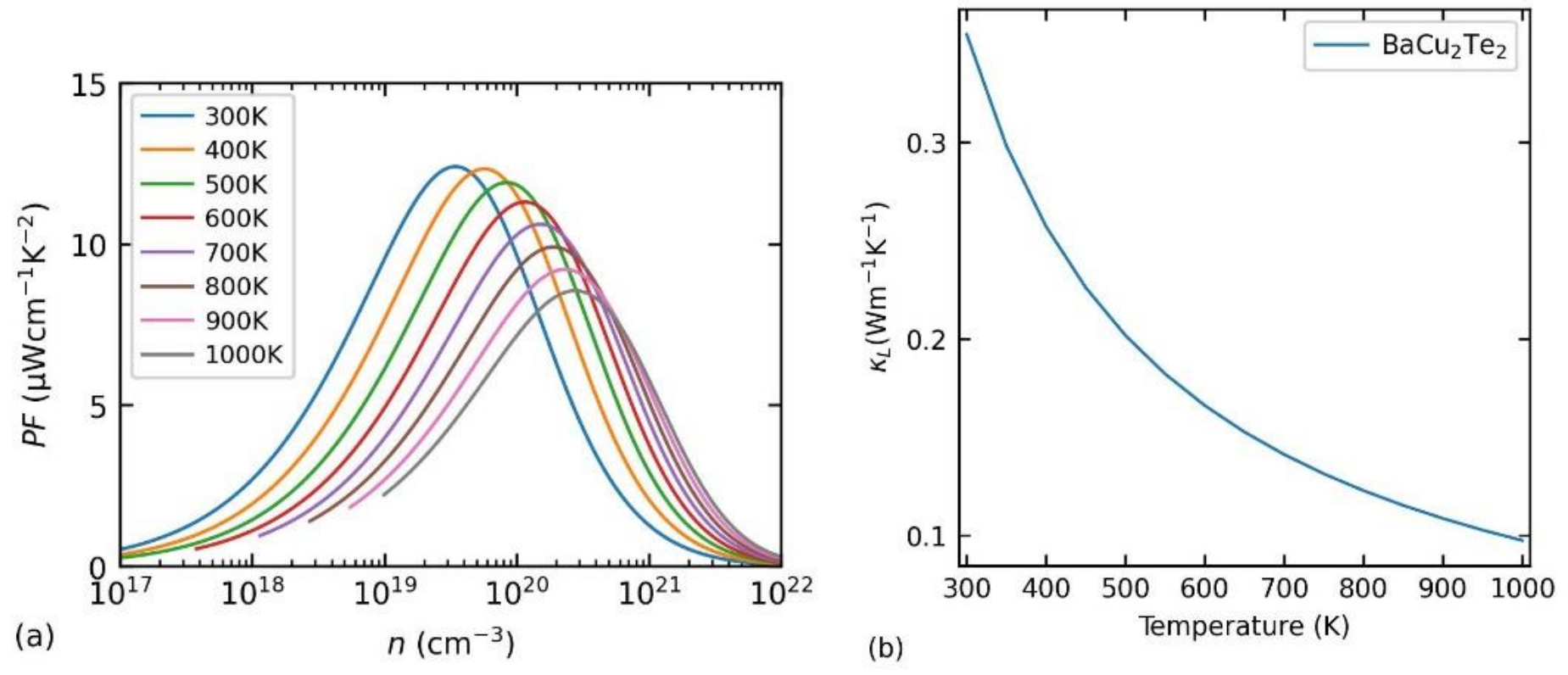

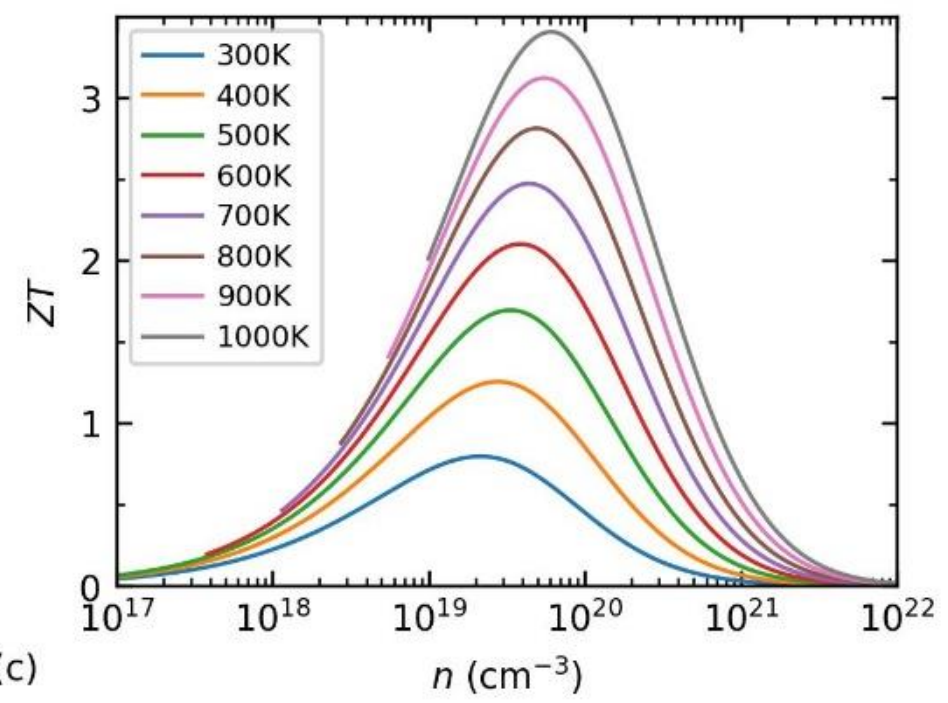

Figure S5. Thermoelectric properties of $BaCu_2Te_2$ for n type transport. (a) power factor, (b) lattice thermal conductivity, and (c) figure of merit.

Table S4. The optimized hyperparameters of the models listed in Table 2

| | Model | Hyperparameters |
|---|---|---|
| N | GBDT | subsample=0.5, min_weight_fraction_leaf=0.003, learning_rate=0.1, max_depth=7, n_estimators=650 |
| | XGB | tree_method='exact', importance_type='gain', scale_pos_weight=1.4404332, n_jobs=4, colsample_bynode=0.7, subsample=0.9, learning_rate=0.08, max_depth=3, n_estimators=100 |
| | AdaB | base_estimator__max_depth=3, learning_rate=0.5, n_estimators=500 |
| | RF | min_samples_split=2, class_weight='balanced_subsample', max_depth=10, max_features=0.2, n_estimators=100 |
| P | GBDT | subsample=0.5, min_weight_fraction_leaf=0.0043, learning_rate=0.1, max_depth=4, n_estimators=600 |
| | XGB | tree_method='exact', importance_type='gain', scale_pos_weight= 2.0954545, n_jobs=4, colsample_bynode=1.0, subsample=0.7, learning_rate=0.09, max_depth=2, n_estimators=500 |
| | AdaB | base_estimator__max_depth=3, learning_rate=0.3, n_estimators=500 |
| | RF | min_samples_split=2, class_weight='balanced_subsample', max_depth=9, max_features=0.4, n_estimators=300 |

Tabel S5. Performance measure on test sets of GBDT algorithm with different input feature vectors. The bold ones are the best for each doping type

| | features | Acc. | Prec. | Recall | F1 | AUC |
|---|---|---|---|---|---|---|
| N | PRDF10 | 0.86 | 0.79 | 0.87 | 0.83 | 0.93 |
| | PRDF16 | 0.84 | 0.76 | 0.90 | 0.82 | 0.94 |
| | PRDF20 | 0.86 | 0.78 | 0.90 | 0.84 | 0.94 |
| | PRDF25 | 0.84 | 0.76 | 0.90 | 0.82 | 0.94 |
| | **GRDF_BOP** | **0.86** | **0.78** | **0.90** | **0.84** | **0.97** |
| | VORONOI | 0.86 | 0.76 | 0.94 | 0.84 | 0.96 |
| P | PRDF10 | 0.89 | 0.90 | 0.75 | 0.82 | 0.96 |
| | **PRDF16** | **0.92** | **0.95** | **0.79** | **0.86** | **0.96** |
| | PRDF20 | 0.89 | 0.9 | 0.75 | 0.82 | 0.95 |
| | PRDF25 | 0.91 | 0.95 | 0.75 | 0.84 | 0.96 |
| | GRDF_BOP | 0.91 | 0.90 | 0.79 | 0.84 | 0.96 |
| | VORONOI | 0.91 | 0.87 | 0.83 | 0.85 | 0.97 |

Table S6. Performance measure on test sets of XGB algorithm with different input feature vectors. The bold ones are the best for each doping type

| | features | Acc. | Prec. | Recall | F1 | AUC |
|---|---|---|---|---|---|---|
| N | PRDF10 | 0.84 | 0.77 | 0.87 | 0.82 | 0.92 |
| | PRDF16 | 0.83 | 0.74 | 0.90 | 0.81 | 0.93 |
| | PRDF20 | 0.84 | 0.76 | 0.90 | 0.82 | 0.94 |
| | PRDF25 | 0.83 | 0.75 | 0.87 | 0.81 | 0.95 |
| | GRDF_BOP | 0.84 | 0.76 | 0.90 | 0.82 | 0.96 |
| | **VORONOI** | **0.88** | **0.81** | **0.94** | **0.87** | **0.95** |
| P | PRDF10 | 0.88 | 0.83 | 0.79 | 0.81 | 0.95 |
| | PRDF16 | 0.86 | 0.84 | 0.67 | 0.74 | 0.94 |
| | PRDF20 | 0.87 | 0.85 | 0.71 | 0.77 | 0.91 |
| | PRDF25 | 0.89 | 0.86 | 0.79 | 0.83 | 0.96 |
| | **GRDF_BOP** | **0.92** | **0.95** | **0.79** | **0.86** | **0.96** |
| | VORONOI | 0.92 | 0.88 | 0.88 | 0.88 | 0.97 |

Table S7. Performance measure on test sets of AdaB algorithm with different input feature vectors. The bold ones are the best for each doping type

| | features | Acc. | Prec. | Recall | F1 | AUC |
|---|---|---|---|---|---|---|
| N | PRDF10 | 0.86 | 0.79 | 0.87 | 0.83 | 0.92 |
| | **PRDF16** | **0.89** | **0.85** | **0.90** | **0.88** | **0.95** |
| | PRDF20 | 0.83 | 0.76 | 0.84 | 0.80 | 0.91 |
| | PRDF25 | 0.86 | 0.79 | 0.87 | 0.83 | 0.93 |
| | GRDF_BOP | 0.89 | 0.83 | 0.94 | 0.88 | 0.96 |
| | VORONOI | 0.87 | 0.77 | 0.97 | 0.86 | 0.97 |
| P | PRDF10 | 0.86 | 0.88 | 0.62 | 0.73 | 0.96 |
| | PRDF16 | 0.89 | 0.86 | 0.79 | 0.83 | 0.93 |
| | PRDF20 | 0.88 | 0.89 | 0.71 | 0.79 | 0.94 |
| | **PRDF25** | **0.89** | **0.9** | **0.75** | **0.83** | **0.93** |
| | GRDF_BOP | 0.88 | 0.89 | 0.71 | 0.79 | 0.94 |
| | VORONOI | 0.86 | 0.81 | 0.71 | 0.76 | 0.93 |

Table S8. Performance measure on test sets of RF algorithm with different input feature vectors. The bold ones are the best for each doping type

| | features | Acc. | Prec. | Recall | F1 | AUC |
|---|---|---|---|---|---|---|
| N | PRDF10 | 0.84 | 0.77 | 0.87 | 0.82 | 0.92 |
| | PRDF16 | 0.84 | 0.76 | 0.90 | 0.82 | 0.91 |
| | PRDF20 | 0.87 | 0.80 | 0.90 | 0.85 | 0.93 |
| | PRDF25 | 0.82 | 0.73 | 0.87 | 0.79 | 0.92 |
| | **GRDF_BOP** | **0.88** | **0.81** | **0.94** | **0.87** | **0.96** |
| | VORONOI | 0.88 | 0.81 | 0.94 | 0.87 | 0.95 |
| P | PRDF10 | 0.86 | 0.84 | 0.67 | 0.74 | 0.91 |
| | PRDF16 | 0.86 | 0.84 | 0.67 | 0.74 | 0.91 |
| | PRDF20 | 0.88 | 0.86 | 0.75 | 0.80 | 0.93 |
| | PRDF25 | 0.88 | 0.86 | 0.75 | 0.80 | 0.94 |
| | **GRDF_BOP** | **0.89** | **0.90** | **0.75** | **0.82** | **0.95** |
| | VORONOI | 0.87 | 0.82 | 0.75 | 0.78 | 0.96 |

Table S9. Comparison of the performance on test sets of GBDT algorithm with input feature vector including band structure descriptors and not (woB)

| | features | Acc. | Prec. | Recall | F1 | AUC |
|---|---|---|---|---|---|---|
| N | GRDF_BOP | 0.86 | 0.78 | 0.90 | 0.84 | 0.97 |
| | GRDF_BOP_woB | 0.80 | 0.77 | 0.74 | 0.75 | 0.88 |
| | VORONOI | 0.86 | 0.76 | 0.94 | 0.84 | 0.96 |
| | VORONOI_woB | 0.75 | 0.70 | 0.68 | 0.69 | 0.84 |
| | PRDF16 | 0.84 | 0.76 | 0.90 | 0.82 | 0.94 |
| | PRDF16_woB | 0.78 | 0.79 | 0.61 | 0.69 | 0.83 |
| P | GRDF_BOP | 0.91 | 0.90 | 0.79 | 0.84 | 0.96 |
| | GRDF_BOP_woB | 0.80 | 0.70 | 0.67 | 0.68 | 0.82 |
| | VORONOI | 0.91 | 0.87 | 0.83 | 0.85 | 0.97 |
| | VORONOI_woB | 0.80 | 0.68 | 0.71 | 0.69 | 0.81 |
| | PRDF16 | 0.92 | 0.95 | 0.79 | 0.86 | 0.96 |
| | PRDF16_woB | 0.83 | 0.74 | 0.71 | 0.72 | 0.80 |

Table S10. Comparison of the performance on test sets of XGB algorithm with input feature vector including band structure descriptors and not (woB)

| | features | Acc. | Prec. | Recall | F1 | AUC |
|---|---|---|---|---|---|---|
| N | GRDF_BOP | 0.84 | 0.76 | 0.90 | 0.82 | 0.96 |
| | GRDF_BOP_woB | 0.80 | 0.77 | 0.74 | 0.75 | 0.88 |
| | VORONOI | 0.88 | 0.81 | 0.94 | 0.87 | 0.95 |
| | VORONOI_woB | 0.75 | 0.70 | 0.68 | 0.69 | 0.84 |
| | PRDF16 | 0.83 | 0.74 | 0.90 | 0.81 | 0.93 |
| | PRDF16_woB | 0.78 | 0.79 | 0.61 | 0.69 | 0.83 |
| P | GRDF_BOP | 0.92 | 0.95 | 0.79 | 0.86 | 0.96 |
| | GRDF_BOP_woB | 0.80 | 0.70 | 0.67 | 0.68 | 0.82 |
| | VORONOI | 0.92 | 0.88 | 0.88 | 0.88 | 0.97 |
| | VORONOI_woB | 0.80 | 0.68 | 0.71 | 0.69 | 0.81 |
| | PRDF16 | 0.86 | 0.84 | 0.67 | 0.74 | 0.94 |
| | PRDF16_woB | 0.83 | 0.74 | 0.71 | 0.72 | 0.80 |

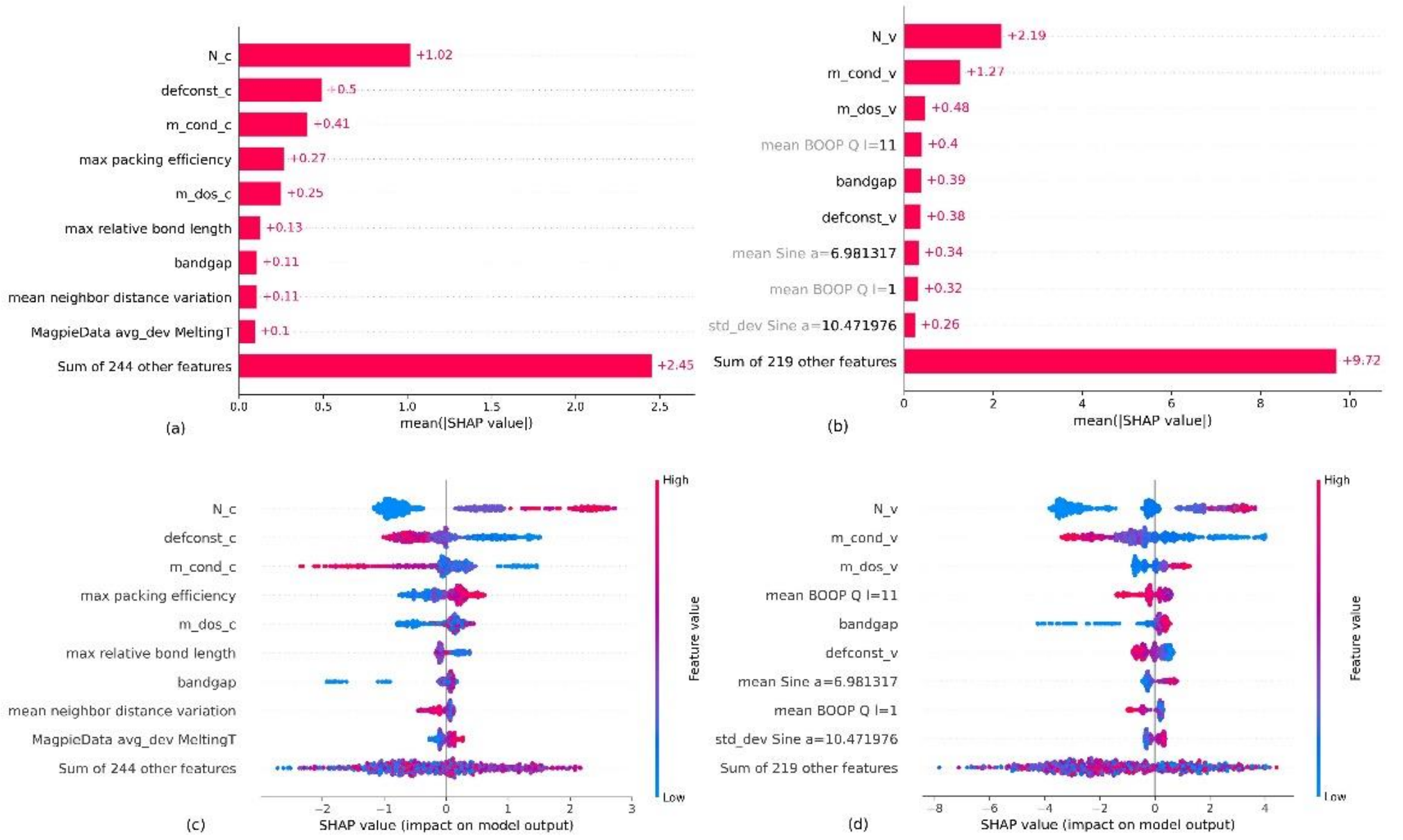

Figure S6. SHAP analysis based on XGB models in Table 2. (a)(c) for n type data, (b)(d) for p type data. (a)(b) bar chart of the average SHAP value magnitude for each input feature. (c)(d) a set of beeswarm plots, where each dot corresponds to a sample. The dot's position on the *x* axis shows the impact that feature has on the model's prediction for that sample. When multiple dots locate at the same x position, they pile up to show density.

Table S11. The optimized hyperparameters of the MEGNet models listed in Table 3

| Hyperparameters | N | P |
|---|---|---|
| Initial learning rate | 0.001 | |
| Final learning rate | 0.0001 | |
| Weight decay | 0.0001 | 0.0056 |
| nblocks | 3 | |
| hidden_layer_sizes_input | (8,4) | |
| hidden_layer_sizes_conv | (8,8,4) | |
| nlayers_set2set | 1 | |
| niters_set2set | 2 | |
| hidden_layer_sizes_output | (8,8) | |
| cutoff | 5.0 | |

Table S12. The optimized hyperparameters of the M3GNet models listed in Table 3

| Hyperparameters | N | P |
|---|---|---|
| Initial learning rate | 0.001 | |
| Final learning rate | 0.0001 | |
| Weight decay | 0.0026 | 0.0048 |
| nblocks | 3 | |
| dim_node_embedding | 32 | 8 |
| dim_edge_embedding | 32 | 8 |
| nlayers_set2set | 1 | |
| niters_set2set | 3 | |
| units | 32 | 8 |
| threebody_cutoff | 5.0 | |

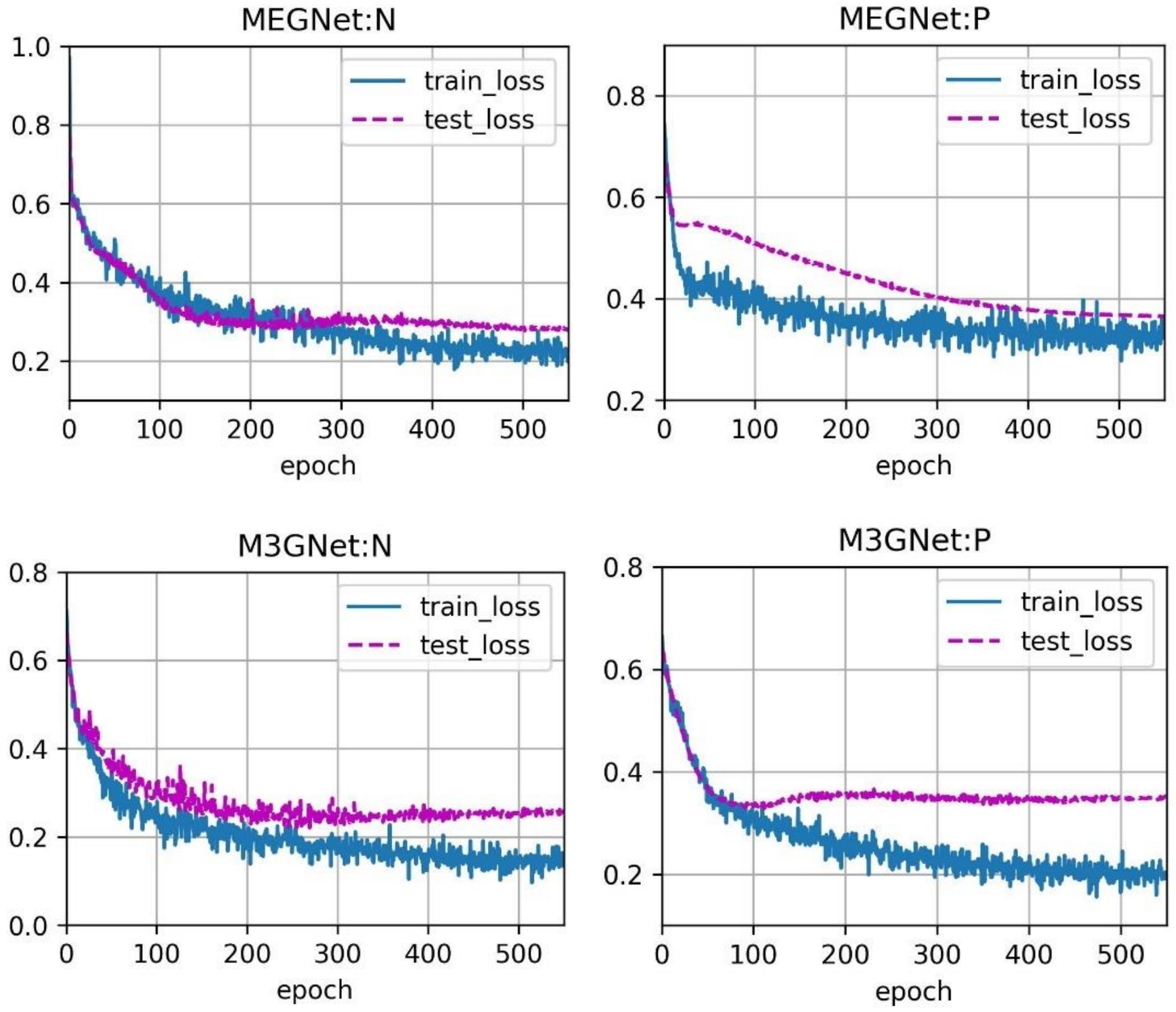

Figure S7. The final training process of the MEGNet and M3GNet models listed in Table 3.

## References

[1] Nielsen, M. D., Ozolins, V., & Heremans, J. P. (2013). Lone pair electrons minimize lattice thermal conductivity. *Energy & Environmental Science*, *6*(2), 570-578.

[2] Zeier, W. G., Zevalkink, A., Gibbs, Z. M., Hautier, G., Kanatzidis, M. G., & Snyder, G. J. (2016). Thinking like a chemist: intuition in thermoelectric materials. *Angewandte Chemie International Edition*, *55*(24), 6826-6841.